T.C.
EGE ÜNİVERSİTESİ
Fen Bilimleri Enstitüsü

# ÇEKİŞMELİ ÜRETİCİ AĞLAR İLE GERÇEKÇİ SAÇ SENTEZİ

**Yüksek Lisans Tezi**

Muhammed PEKTAŞ

Bilgisayar Mühendisliği Anabilim Dalı

T.C.
EGE ÜNİVERSİTESİ
Fen Bilimleri Enstitüsü

# ÇEKİŞMELİ ÜRETİCİ AĞLAR İLE GERÇEKÇİ SAÇ SENTEZİ

Muhammed PEKTAŞ

Danışman: Prof. Dr. Aybars UĞUR

Bilgisayar Mühendisliği Anabilim Dalı
Bilgisayar Mühendisliği Yüksek Lisans Programı

**EGE ÜNİVERSİTESİ FEN BİLİMLERİ ENSTİTÜSÜ**

**ETİK KURALLARA UYGUNLUK BEYANI**

EÜ Lisansüstü Eğitim ve Öğretim Yönetmeliğinin ilgili hükümleri uyarınca Yüksek Lisans Tezi olarak sunduğum "**ÇEKİŞMELİ ÜRETİCİ AĞLAR İLE GERÇEKÇİ SAÇ SENTEZİ** " başlıklı bu tezin kendi çalışmam olduğunu, sunduğum tüm sonuç, doküman, bilgi ve belgeleri bizzat ve bu tez çalışması kapsamında elde ettiğimi, bu tez çalışmasıyla elde edilmeyen bütün bilgi ve yorumlara atıf yaptığımı ve bunları kaynaklar listesinde usulüne uygun olarak verdiğimi, tez çalışması ve yazımı sırasında patent ve telif haklarını ihlal edici bir davranışımın olmadığını, bu tezin herhangi bir bölümünü bu üniversite veya diğer bir üniversitede başka bir tez çalışması içinde sunmadığımı, bu tezin planlanmasından yazımına kadar bütün safhalarda bilimsel etik kurallarına uygun olarak davrandığımı ve aksinin ortaya çıkması durumunda her türlü yasal sonucu kabul edeceğimi beyan ederim.

28 / 01 / 2022

Muhammed PEKTAŞ

# ÖZET

## ÇEKİŞMELİ ÜRETİCİ AĞLAR İLE GERÇEKÇİ SAÇ SENTEZİ

PEKTAŞ, Muhammed

Yüksek Lisans Tezi, Bilgisayar Mühendisliği Anabilim Dalı
Tez Danışmanı: Prof. Dr. Aybars UĞUR
Ocak 2022, 54 sayfa

Üretici modelleme alanında elde edilen son yıllardaki başarılar, bu konudaki çalışmaları hızlandırmış ve araştırmacıların ilgisini çekmiştir. Bu başarıda en çok payı olan yöntemlerden bir tanesi de çekişmeli üretici ağlar'dır. Çekişmeli üretici ağlar, sanal gerçeklik (VR), artırılmış gerçeklik (AR), süper çözünürlük (super-resolution), görüntü iyileştirme (image enhancement) gibi konularda bir çok uygulama alanına sahiptir.

Saç sentezleme, stil transferi ve şekil düzenleme görevleri, derin öğrenme ve üretici modellemedeki son ilerlemelere rağmen saçın karmaşık doğası nedeniyle hala çözülmeyi bekleyen zorlukları içerisinde barındırmaktadır. Bu problemin çözümü için literatürde önerilen yöntemler genellikle görseller üzerinde yüksek kaliteli saç düzenlemeleri yapmak üzerine odaklanmıştır. Bu tez kapsamında saç sentezi probleminin çözümü için bir çekişmeli üretici ağ yöntemi önerilmiştir. Bu yöntem geliştirilirken, literatürdeki en iyi yöntemler ile yarışır görsel çıktılar elde edebilmenin yanısıra, gerçek zamanlı saç sentezi yapılabilmesi de amaçlanmıştır.

Önerilen yöntem, FFHQ veri kümesi ile eğitilmiş ve devamında ise saç stili transferi ve saç yeniden yapılandırma görevlerindeki sonuçları değerlendirilmiştir. Bu görevlerde elde edilen sonuçlar ve yöntemin işletim süresi literatürdeki en iyi yöntemlerden biri olan MichiGAN ile karşılaştırılmıştır. Karşılaştırma 128x128 çözünürlüğünde gerçekleştirilmiştir. Karşılaştırma sonucunda önerilen yöntemin, gerçekçi saç sentezi açısından MichiGAN ile yarışır sonuçlar elde ettiği, işletim süresi açısından daha iyi performans sergilediği gösterilmiştir.

# ABSTRACT

## REALISTIC HAIR SYNTHESIS WITH GENERATIVE ADVERSARIAL NETWORKS

PEKTAŞ, Muhammed

MSc in Computer Engineering
Supervisor: Prof. Dr. Aybars UĞUR

Recent successes in generative modeling have accelerated studies on this subject and attracted the attention of researchers. One of the most important methods used to achieve this success is Generative Adversarial Networks (GANs). It has many application areas such as; virtual reality (VR), augmented reality (AR), super resolution, image enhancement.

Despite the recent advances in hair synthesis and style transfer using deep learning and generative modelling, due to the complex nature of hair still contains unsolved challenges. The methods proposed in the literature to solve this problem generally focus on making high-quality hair edits on images. In this thesis, a generative adversarial network method is proposed to solve the hair synthesis problem. While developing this method, it is aimed to achieve real-time hair synthesis while achieving visual outputs that compete with the best methods in the literature.

The proposed method was trained with the FFHQ dataset and then its results in hair style transfer and hair reconstruction tasks were evaluated. The results obtained in these tasks and the operating time of the method were compared with MichiGAN, one of the best methods in the literature. The comparison was made at a resolution of 128x128. As a result of the comparison, it has been shown that the proposed method achieves competitive results with MichiGAN in terms of realistic hair synthesis, and performs better in terms of operating time.

# ÖNSÖZ

Saç sentezi problemi günlük hayat içerisinde birçok uygulama alanına sahiptir ve bir çok yeni uygulama alanı için de gelecek vaad etmektedir. Bu alanların başında sanal gerçeklik ve artırılmış gerçeklik uygulamaları olmakla birlikte çeşitli sosyal medya uygulamaları ve hatta kayıp ve aranan kişilerin muhtemel görünümlerini tahmin etmek gibi problemleri içeren uygulamalar da yer almaktadır.

Bu tezde, görece yeni bir alan olan çekişmeli üretici ağlar ile saç sentezi problemi üzerinde çalışılmıştır. Tez çalışması kapsamında literatürdeki en iyi yöntemler ile yarışır kalitede saç sentezi üretimi yapabilen ve aynı zamanda işletim süresi olarak verimli bir yöntem üzerinde çalışılmıştır. Önerilen yöntem saç yeniden yapılandırma, saç stili transferi görevlerinde test edilmiştir. Testler sonucunda literatürdeki en iyi yöntemler ile yarışır sonuçlar elde edilmiştir.

İZMİR

28 / 01 / 2022 Muhammed PEKTAŞ

# İÇİNDEKİLER

Sayfa

# İÇİNDEKİLER (devam)

Sayfa

# İÇİNDEKİLER (devam)

Sayfa

# ŞEKİLLER DİZİNİ

Şekil Sayfa

## ŞEKİLLER DİZİNİ (devam)

<u>Şekil</u> <u>Sayfa</u>

# TABLOLAR DİZİNİ

Tablo Sayfa

# SİMGELER VE KISALTMALAR DİZİNİ

| Simgeler | Açıklama |
|---|---|
| b | Sabit değer |
| W | Ağırlıklar |
| f | Fonksiyon |
| K | Filtre |
| M | Özellik haritası |
| X | Giriş değişkeni |
| Y | Çıkış değişkeni |
| s | Stil vektörü |
| $M_L$ | L. katmanın maske girdisi |
| $I_{bg}$ | Girdi arkaplanı |
| F | Üretilen görüntü |
| FM | Özellik haritası |
| $K_n$ | n. evrişim işlemi filtresi |
| η | Öğrenme oranı |

Kısaltmalar

| | |
|---|---|
| AdaIN | Adaptive Instance Normalization |
| ANN | Artificial Neural Networks |
| AR | Augmented Reality |

## SİMGELER VE KISALTMALAR DİZİNİ (devam)

| Kısaltmalar | Açıklama |
|---|---|
| CNNs | Convolutional Neural Networks |
| DL | Deep Learning |
| FFHQ | Flicker Face High Quality Dataset |
| ML | Machine Learning |
| VR | Virtual Reality |
| 3D | 3-Dimensional |

# 1. GİRİŞ

Tanımlayıcı modelleme (descriptive modelling) ve üretici modelleme (generative modelling) makine öğrenmesinde kullanılan iki temel yaklaşımdır. Tanımlayıcı modellemede model, veri kümesini girdi olarak alır ve veriye dair bir çıkarımda bulunur. Üretici modellemede ise tam tersi bir durum söz konusudur. Bu yaklaşımda çıktılardan verinin kendisi modellenmeye çalışılır. Bu tezin teorik altyapısını oluşturan üretici modellemenin bir çok kullanım alanı söz konusudur. Sanal Gerçeklik (VR) ve Artırılmış Gerçeklik (AR) uygulamalarında, 3D modelleme, görüntü tamamlama, görüntü iyileştirme, sanatsal çıktılar elde etmek gibi bir çok uygulama alanında kullanılmaktadır. Bu tez kapsamında üretici modelleme yaklaşımı ile gerçekçi saç üretimi üzerinde çalışılmıştır. Görece yeni bir alan olmasından dolayı, bu alan dar bir literatüre sahiptir. Bu tez kapsamında üzerinde çalışılan yöntem, literatürdeki en iyi yöntemlerle yarışır sonuçlar vermektedir.

Görüntülerde gerçekçi saç sentezleme işleminin kullanılabileceği bir çok önemli alan vardır. Örneğin bu tez kapsamında yapılan çalışma ile kolluk kuvvetleri, saç rengi ve şeklini değiştirerek kimliği gizlenen kayıp veya suçlu bireylerin olası durumları için farklı görüntüler elde edebilirler ya da saçının görünümünde ve tarzında değişiklik yapmak isteyen kişiler, fiziksel görünümünde uzun süreli değişiklik yapmadan olası durumlarını görüp önceden bir değerlendirme yapabilirler. Bir başka örnek ise çeşitli sosyal medya uygulamalarında kişiler saçları üzerinde değişiklikler yaparak istedikleri görselleri elde edebilirler (Olszewski et al., 2020). Ayrıca farklı saç görünümleri ve tarzları kişinin tanınabilirliği ile ilişkili olduğundan, bu alanlarda değişiklik yapabilme yeteneği, görsel verilerin anonimleştirilmesi süreçlerine de katkı sağlayabilir.

Saç sentezleme problemi tek başına ele alındığında görsel üzerinde sadece saçın değiştiği görsellerin elde edilmesi üzerine çalışılan bir problemdir. Bu problem içerisinde bir çok zorlu alt problem içerir. İnsan yüzünün diğer bölümlerinin aksine kendine özgü geometrisi ve materyali saçın analiz, temsil ve üretim işlemlerini zorlaştırmaktadır. Saç, biraz daha yakından incelenip, şekil ve görünümü ayrı değerlendirildiğinde görsel kalite için birkaç temel husus tanımlanabilir. Şekil açısından karışık saç maskesi sınırları ve farklı opaklık seviyeleri hassas saç şekli kontrolü ve arkaplan ile saçın kusursuz kesişimi konusunda zorluklara neden olabilir. Görünüm açısından ise, saç liflerinin karmaşık doğası, ipliksi yapıların farklı yön ve görünümlerindeki farklılıklar ve bunların arasındaki uyumu sağlama gerekliliği, görünüm açısından da saç üretim, temsil ve analizini zorlaştıran faktörlerdir (Tan et al., 2020). Tüm bu zorluklar olmasına karşın, bu problemlerin çözümü için bir çok çalışma yapılmış ve farklı seviyelerde başarılar elde edilmiştir. Bu çalışmada verilen farklı referans saç görünümlerine sahip yeni saç üretimi yapabilen bir çerçeve üzerinde çalışılmıştır.

Bu çalışmada, üzerinde çalışılacak saç sentezi problemi için stil transferi yapılmış gerçek değerler (ground truth) bulunmadığı için tercih edilen eğitim yöntemine uygun olacak şekilde yüz ve saç bölgelerini içeren çok bilinen veri kümeleri kullanılmıştır. Kullanılan veri kümeleri ile geliştirilen yöntemin eğitim ve değerlendirme deneyleri yapılmıştır. Değerlendirmeler için hem niceliksel (quantitative) hemde niteliksel (qualitative) yöntemler kullanılmıştır. Tez çalışması kapsamında, ifade edilen problem için literatürdeki en iyi sonuçlarla yarışır sonuçlar elde edilirken aynı zamanda çalışma süresinin de azaltılması hedeflenmiştir.

Bu çalışmanın literatüre temel katkısı, literatürdeki yüksek başarı elde edebilen yöntemler ile yarışır sonuçlar elde ederken diğer yöntemlerden daha az çalışma zamanı süresine sahip olan yeni bir yöntemin önerilmiş olmasıdır. Geliştirilen yöntem, saç sentezinin videolar üzerinde gerçek zamanlı gerçekleştirilmesinin de önünü açmaktadır.

Tez metninin sonraki bölümleri şu şekilde özetlenebilir:

İkinci bölümde derin öğrenme, yapay sinir ağları (ANN), evrişimli sinir ağları (CNN), eniyileme konuları temel alt başlıklarıyla birlikte açıklanmaktadır.

Tezin üçüncü bölümünde derin öğrenme teknikleri tabanlı Üretici modelleme çalışmaları üzerinde durulmaktadır. Üretici modelleme konusunda sadece tez ile ilgili önemli başlıklara değinilmektedir.

Dördüncü bölümde, saç sentezi problemi hakkında yapılan çalışmalara değinilecektir. Bu çalışmaların yaklaşımları, kullanılan veri kümeleri ve değerlendirme yöntemleri incelenmektedir.

Beşinci bölümde çekişmeli üretici ağlar ile saç sentezi konusunda önerilen yöntem genel çerçevesi ve tüm alt modülleri ile birlikte açıklanmıştır. Yöntemde kullanılan tüm kayıp fonksiyonları kullanım nedenleriyle birlikte paylaşılmıştır ve yöntemin genel kayıp fonksiyonu tanımlanmıştır.

Altıncı bölümde, ilk olarak önerilen yöntemin eğitimi için kullanılan veri kümeleri ve hazırlanışları ifade edilmiştir. Ön hazırlık aşaması tamamlanmış veriler üzerinde, hazırlanan deney ortamında önerilen yöntem kullanılarak saç stil transferi ve saç yeniden yapılandırma görevleri için deneyler yapılıp elde edilen sonuçlar paylaşılmıştır ve literatür ile karşılaştırmalar yapılmıştır.

Tezin son bölümünde ise literatüre yapılan katkılar ile birlikte yapılan çalışmanın sonuçlar ve gelecek çalışma önerileri sunulmaktadır.

# 2. DERİN ÖĞRENME

Bu bölümde Derin öğrenmenin temel kavramları kısaca anlatılmaktadır.

## 2.1 Yapay Sinir Ağları

Derin öğrenme çok geniş bir literatüre sahiptir. Başta bilgisayarlı görü (computer vision) olmak üzere bir çok alanda önemli etkilere sahiptir. Son yılların en popüler ve verimli araştırma alanı olarak görülmektedir. Özellikle derin öğrenme ile gösterimli öğrenme yöntemleri kullanılarak bazı görevlerde insan başarısıyla benzer veya daha başarılı sonuçlar elde edilmiştir (Goh et al., 2020; Stidham et al., 2019).

Derin öğrenme ilhamını beyindeki sinir sisteminden alır. Nöron, sinir sisteminin temel fonksiyonel birimidir. Başlıca görevi bilgi transferini gerçekleştirmesidir. Nöronlar kendilerine gelen bilgiyi (sinyali) toplar, gerekirse aktivasyon seviyesinde değişime sebep olur ve çıktıyı diğer nöronlara aktarırlar. Biyolojik sinir ağlarından esinlenilerek yapay sinir ağları önerilmiştir. Standart bir yapay sinir ağı bir veya birden çok birbirine bağlı temel iş biriminden oluşur. Bu iş birimi yapay nöron olarak adlandırılır. Yapay nöronlar girdi olarak aldığı bilgiyi belirli bir ağırlık ile çarpar ve ek bir girdi ile toplayarak sonuca aktarır. Oluşan bu çıktı, uygulanmak istendiği takdirde nöronun aktiflik durumuna karar verecek bir aktivasyon fonksiyonundan geçirilir. Bu aşamada elde edilen çıktı başka nöronların girdisi veya nihai sonuç olabilir. Her biyolojik nöron, dendrit adı verilen sinyal toplayıcı uçlardan girdi sinyalini alır, hücre çekirdeğinde gerekli işlemlerini yapar ve onu akson boyunca taşıyarak çıktı sinyalini üretir. Biyolojik nörondan esinlenilen bu işlemin matematiksel olarak modellenmesi Şekil 2.1'de gösterilmiştir.

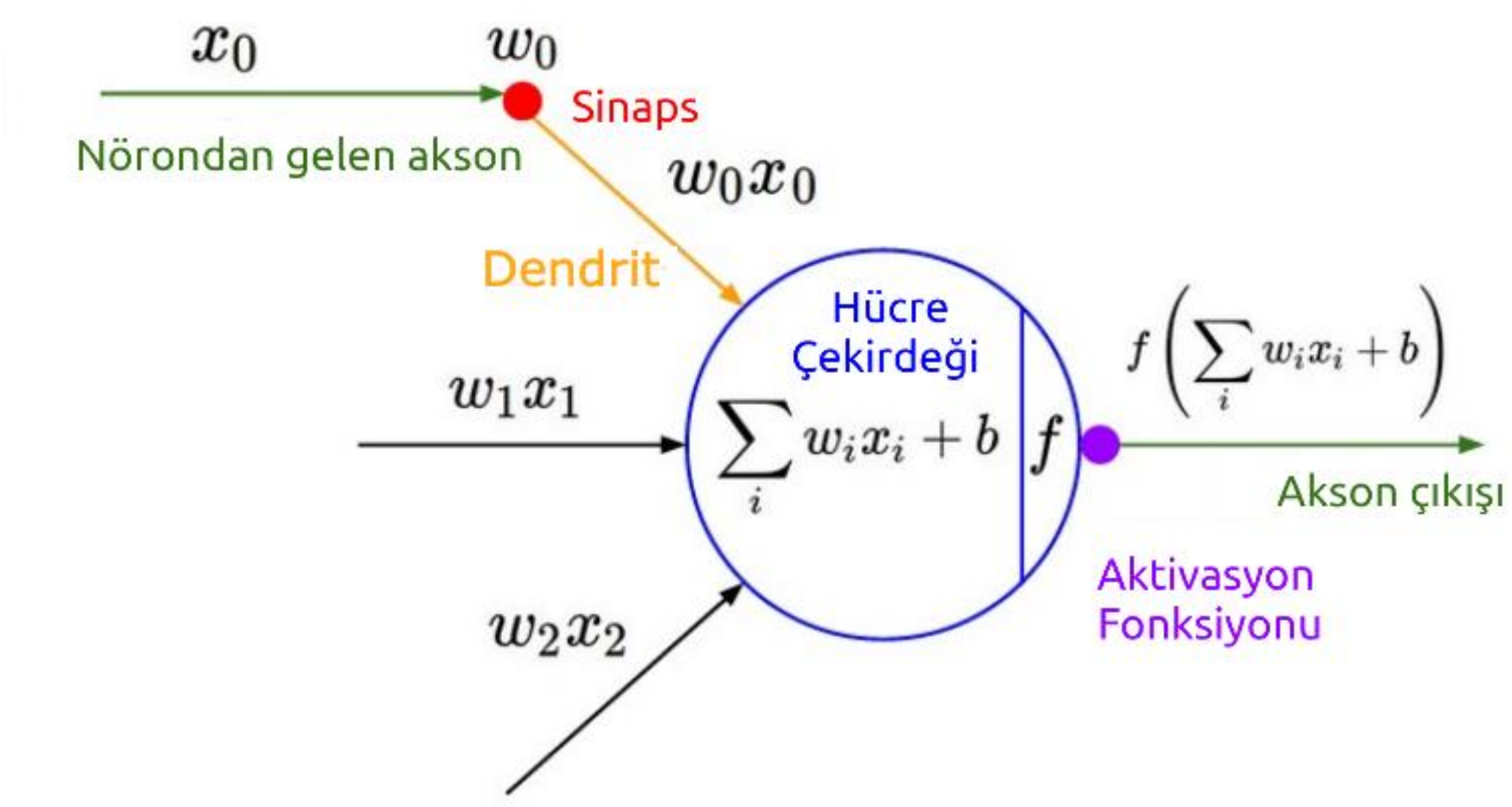

Şekil 2.1. Biyolojik nöronun matematiksel modellemesi (University of Stanford, n.d.)

$(x_0)$ girdi, $(w_0)$ girdinin ağırlığı, $(b)$ ek girdi, $(f)$ aktivasyon fonksiyonu olmak üzere bir yapay nöronun çıktısı, tüm girdiler ağırlıkları ile çarpılarak ek değerle toplanılıp bir aktivasyon fonksiyonundan geçirilerek elde edilir. Bir nöron için bu işlem denklem 2.1 ile ifade edilebilir.

$$y = f\left(\sum (w_i * x_i + b)\right) \qquad \textbf{2.1}$$

Buradaki aktivasyon fonksiyonu, nöronun aktivasyon seviyesini belirler. Aktivasyon fonksiyonu olmadan denklemin kendisi doğrusal bir denklem olduğundan doğrusal olmayan bir problemi modelleyebilmek için genellikle doğrusal olmayan aktivasyon fonksiyonları kullanılır. Şekil 2.2’de bazı örnek klasik aktivasyon fonksiyonları gösterilmiştir.

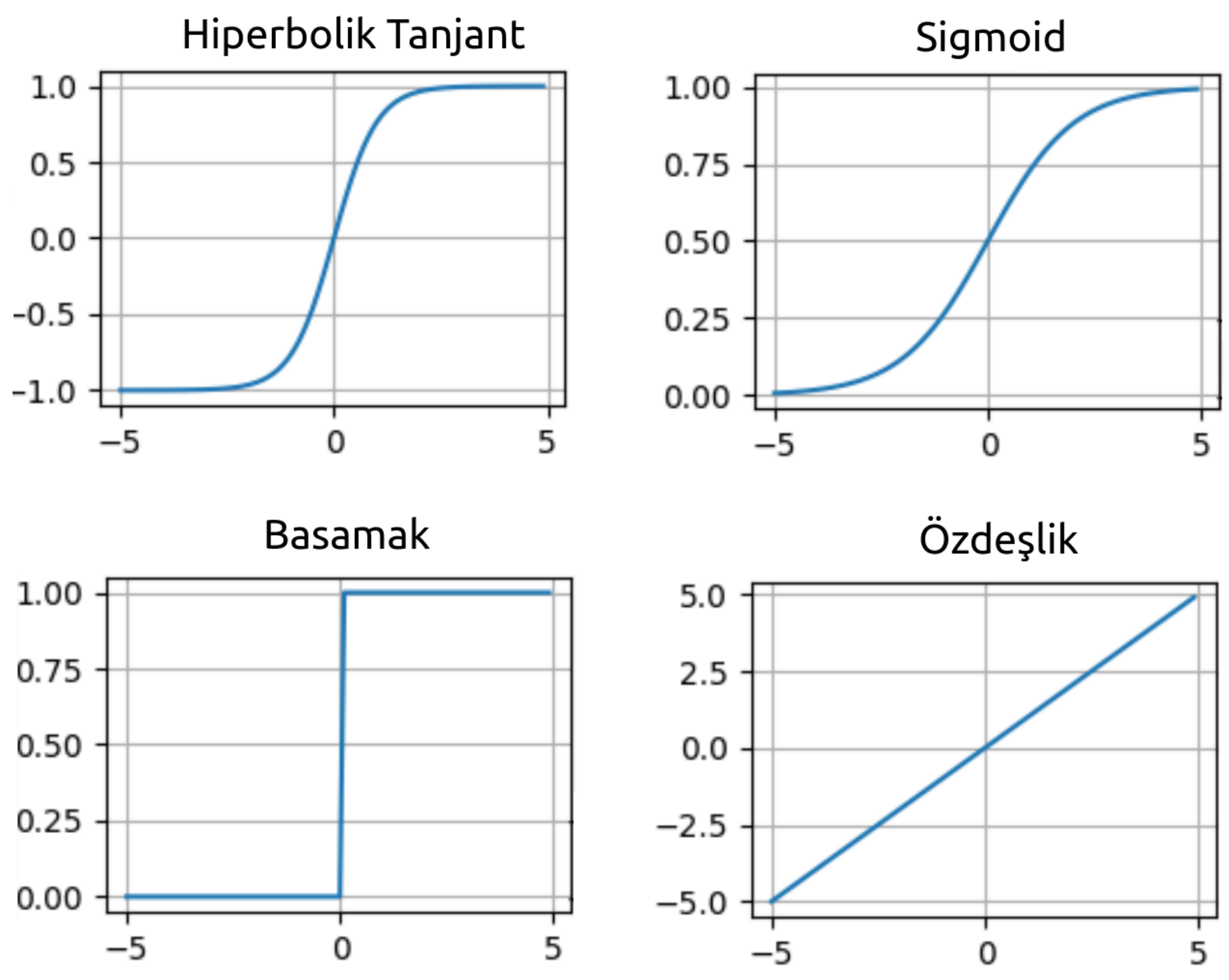

Şekil 2.2. Bazı klasik aktivasyon fonksiyonlarının girdisi [-5,5] aralığında iken sahip oldukları değerleri (Apicella et al., 2021).

Sigmoid fonksiyonu [0,1] aralığında değerler alır. Değer aldığı aralık sebebiyle ara katmanlarda genellikle aktivasyonları maskeleyici çıktılar oluşturmak veya çıktı tahminlerini olasılık olarak ifade etmek için kullanılır.

Hiperbolik tanjant fonksiyonu [-1,1] aralığında değer alır. Negatif değerlerin ve sıfıra yakın değerlerinde kendine uygun şekilde çıktıya aktarılması hiperbolik tanjantın avantajları arasında sayılabilir.

Özdeşlik fonksiyonu girdi olarak aldığı değeri doğrudan çıktıya aktarır. Doğrusal bir fonksiyon olduğundan doğrusal olmayan problemlerin çözümünde kullanılamaz. Güncel yapay sinir ağları modellerinde sık kullanılmayan bir aktivasyon fonksiyonudur.

Basamak fonksiyonu genellikle ikili sınıflandırma problemlerinin çözümünde son katmanlarda kullanılır. Türev öğrenebilme yeteneği olmadığından ara katmanlarda kullanımı tavsiye edilmez.

Belirtilen girdiler, ağırlıklar, ek girdiler ve aktivasyon fonksiyonları ışığında bir yapay nöronun çıktısı başka bir yapay nöronun girdisi olabilir böylece bir sinir ağı oluştururlar. Oluşturulan sinir ağının kapasitesi arttıkça daha büyük bir hipotez uzayına sahip olur ve daha zorlu problemlerin modellenmesine imkan verir. Hipotez uzayı bir modelin ifade edebileceği çözüm kümesidir (Goodfellow et al., 2016).

## 2.2 Evrişimsel Sinir Ağları

Geleneksel makine öğrenmesi algoritmalarında problemin çözümü için gerekli özellik çıkarımı araştırmacılar tarafından yapılmaktadır. Bu özellik çıkarma işlemi ayrıca alanda uzmanlık bilgisi de gerektirmektedir ancak CNN (LeCun et al., 1989) uygulamalarında özellik çıkarımı öğrenme sırasında ara katmanlarda elde edilebilmektedir. Bu durum daha az uzman bilgisi gerektirmesi, özellik çıkarımı sırasındaki zaman kazanılması anlamında önemli bir kazançtır. Bu nedenle CNN, örüntü tanıma ile ilgili alanlarda son yıllarda önemli bir etkiye sahiptir.

Görüntü verileri uzamsal ilişkileri olan görsellerdir ve CNN bu özelliği kullanır. CNN de filtreler kümesi öğrenilir. $[K_1 \ldots \ldots K_n]$ filtreler kümesi, (n) filtre indeksi olmak üzere filtreler, (X) girdi görüntüsü ile evrişim işleminden geçirilir ve sonucunda (FM) özellik haritaları oluşturulur. Bu işlem denklem 2.2 ile ifade edilebilir.

$$FM_n = K_n \otimes X \qquad \textbf{2.2}$$

Evrişim işleminde kullanılan filtre uygulandığı görüntüyle uyumluysa güçlü, uyumsuz ise zayıf çıktılar üretir. Her bir filtrelerin ağırlık değerleri paylaşıldığından tam bağlı yapay sinir ağlarına göre daha az parametre içerirler. Evrişim işlemi için filtrelerin boyutu (kernel size), adım sayısı (stride), doldurma (padding) işlemi gibi özelliklerin ayarlanması gerekir. Elde edilmek istenilen özellik haritasının içeriği ve büyüklüğünü bu değerle ile ayarlanabilir. Girdi büyüklüğü $(n)$, çıktının büyüklüğü $(n')$, doldurma değeri $(p)$, uygulanan filtre büyüklüğü ($f$), adım değeri ($s$) olmak üzere evirişi işlemi sonucunda oluşan çıktının boyutları denklem 2.3 ile ifade edilebilir.

$$n' = \frac{(n + 2p - f)}{s} - 1 \qquad \textbf{2.3}$$

Evrişim işlemi sırasında kullanılan filtre büyüklüğü, adım sayısı gibi değişkenler çıktıyı etkilemektedir. Örneğin, büyük filtreler özellik çıkarırken daha çok sayıda komşu piksel'i değerlendirir, adım sayısı büyüklüğü filtrelerin resme uygulanma sayısını değiştirebilir. Bu ve benzeri parametreler evrişim işlemi sonucunda elde edilecek özellik haritasını etkilemektedir. Örnek bir evrişim işlemi şekil 2.3 te gösterilmektedir.

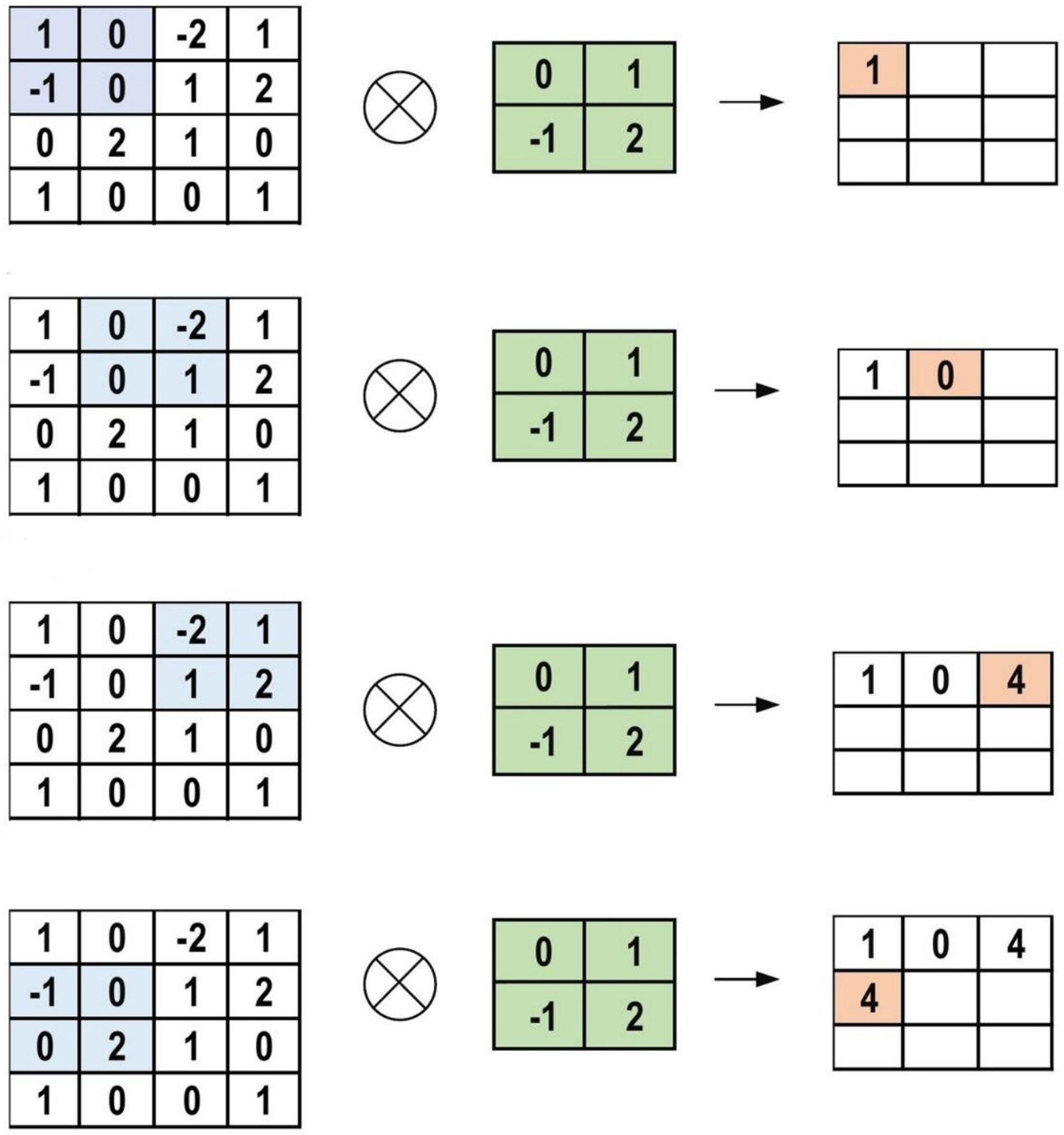

Şekil 2.3. Örnek bir evrişim işleminin bir kaç adımlık gösterimi (Alzubaidi et al., 2021)

CNN’in bir başka özelliği ise yapay sinir ağlarındaki parametre sayısını azaltmasıdır. Bu sayede araştırmacıların daha karmaşık problemleri çözebilmek için daha büyük modeller geliştirmesinin önü açılmıştır. CNN'in bir diğer önemli yönü ise katmanlar ilerledikçe daha soyut özellikler öğrenebilmesidir. Örneğin görsel girdilerle çalışılan bir problemde, ilk katmanda kenar ve köşe özellikleri ikinci katmanda ise basit şekiller olacak şekilde öğrenmeye etki eden özellikler öğrenilir (Albawi et al., 2017). Şekil 2.4 te öğrenilen özelliklerin temsili gösterilmiştir.

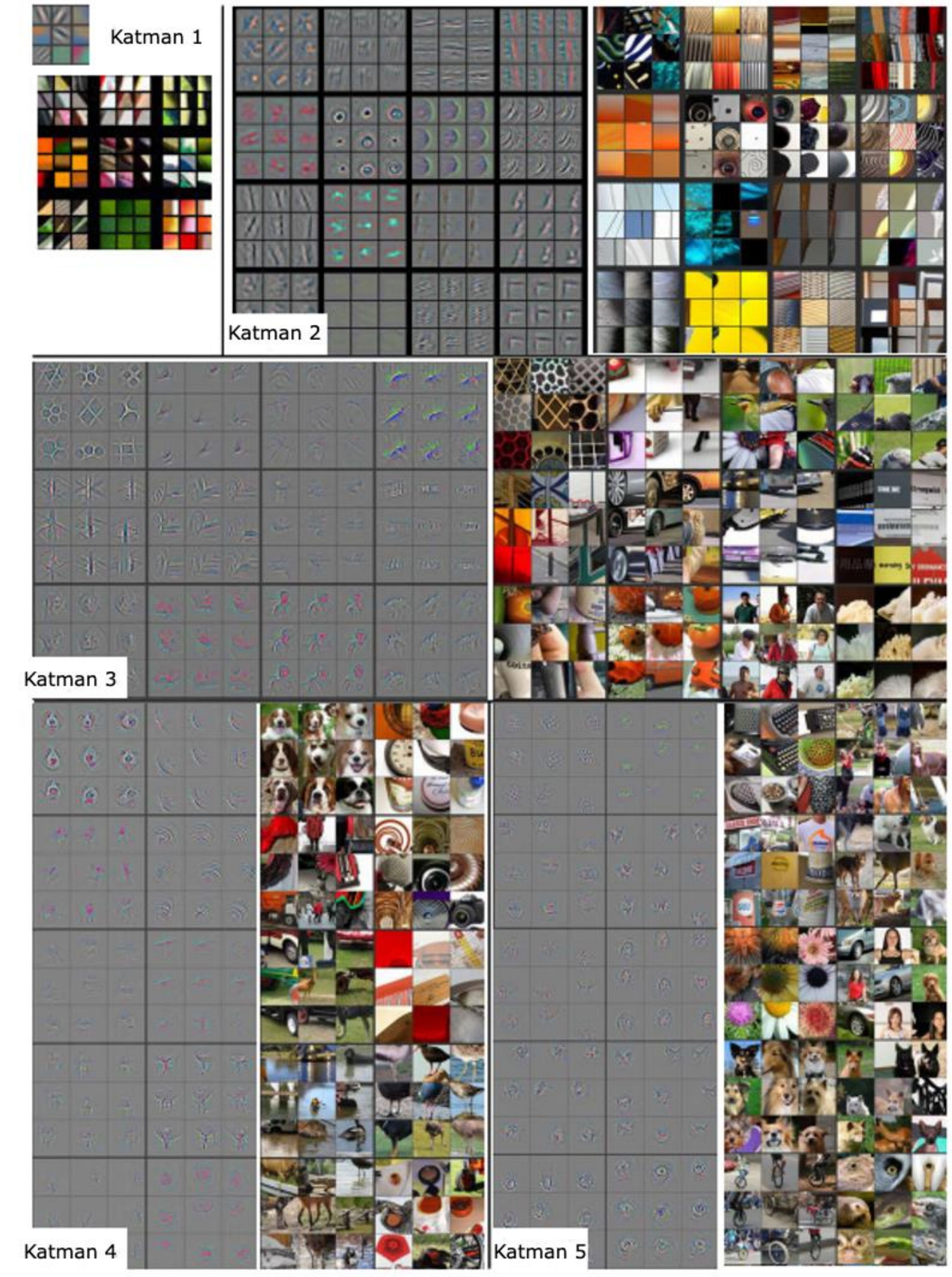

Şekil 2.4. Katmanlar ve öğrenilen özellik örnekleri (Zeiler et al., 2014)

## 2.3 Eniyileme

Yapay sinir ağlarında öğrenme işlemi temelde, ağırlıklar ve ek birim değerlerinin bir veri kümesi üzerinde en az hatayı verecek şekilde değiştirilmesi işlemidir. Bu işlem aslında bir en iyileme (optimization) problemidir. Çoğu en iyileme yöntemi girdileri değiştirerek ilgili fonksiyonunun değerini en küçültmeyi (minimization) amaçlar. Eniyileme işlemi uygulamak istediğimiz fonksiyon genellikle amaç yada kayıp fonksiyonu olarak adlandırılır. Yapay sinir ağlarında en sık kullanılan eniyileme yöntemi gradyan inişi yöntemidir (Ruder et al., 2016).

### 2.3.1 Gradyan iniş

Gradyan iniş algoritması yapay sinir ağlarını optimize etmede kullanılan en bilinir algoritmadır ve son yıllarda da bilinirliğini giderek artmaktadır. Keras (Chollet et al., 2015), Pytorch (Paszke et al., 2019) gibi birçok iyi bilinen derin öğrenme kütüphanelerinde birçok farklı versiyonunun uygulamaları bulunmaktadır. Bu algoritma genellikle kara kutu bir eniyileme yöntemi olarak kullanılır.

Gradyan iniş algoritması, ( $\theta \in R^d$) model parametreleri olmak üzere $J(\theta)$ amaç fonksiyonunu (örneğin bölüm 2.4 deki fonksiyonlar) optimize eder. Optimizasyon sürecinde amaç fonksiyonunun parametrelere göre gradyanı ($\Delta\theta J(\theta)$) hesaplanır. Gradyan, fonksiyonun türevini ifade etmektedir. Parametreler gradyanların tam tersi yönde güncellenir. Böylelikle iç bükey olarak modellenmiş problem uzayında minimum değere yaklaşılmış olur. Çözüm uzayının morfolojik gösteriminde aşağıya doğru ilerlendiğinden gradyan iniş algoritması olarak adlandırılır. Gradyan iniş algoritması bazı parametrelere sahiptir. Bunlardan birisi öğrenme oranı ($\eta$) olarak bilinir. Bu oran yerel minimuma yaklaşılırken atılacak adım büyüklüğünü etkiler. Diğer bir deyişle amaç fonksiyonunun oluşturduğu uzayda bir iniş karakteristiği gösterilir (Ruder et al., 2016). Bu algoritmanın yığın gradyan inişi, Stokastik gradyan inişi, Mini-yığın gradyan inişi olmak üzere temelde 3 farklı çeşidi vardır.

### 2.3.2 Yığın gradyan iniş

Yığın gradyan iniş vanilya gradyan iniş olarak bilinir. Bu versiyonda veri kümesindeki tüm elemanların amaç fonksiyonu çıktısı ile yapay sinir ağının parametreleri arasındaki gradyan hesaplanır ve denklem 2.4 ile güncellenir.

$$\theta = \theta - \eta . \Delta\theta J(\theta) \quad \textbf{2.4}$$

Bu yöntemde tek bir güncelleme işlemi için tüm datasetin gradyanları hesaplandığından diğer versiyonlara göre yavaştır ve daha çok bellek ihtiyacına sahiptir.

### 2.3.3 Stokastik gradyan iniş

Stokastik gradyan inişte yığın gradyan inişinin aksine güncelleme işlemi tüm veri seti için bir kere değil, seçilmiş her rastgele örnek için gerçekleştirilir. $(X_i)$ veri kümesinde bir eleman, $(y_i)$ çıktısı olmak üzere güncelleme işlemi denklem 2.5 ile hesaplanır.

$$\theta = \theta - \eta \cdot \nabla \theta J(\theta; X_i ; y_i) \quad \textbf{2.5}$$

Bu işlemde güncelleme için tüm datasetteki gradyanları hesaplamak gerekmediğinden bellek ve hız olarak yığın gradyan iniş yöntemine göre daha verimli bir yoldur ancak en iyileme süresince varyansı yüksek bir yol izleyebilir.

### 2.3.4 Mini-yığın gradyan iniş

Mini-yığın gradyan inişi yönteminde ise eniyileme yolunda düşük varyans ve aynı zamanda hızlı bir optimizasyon yapabilmek adına veri kümesinden küçük bir alt küme seçilir ve bu kümedeki örnekler için gradyanlar hesaplanarak parametreler güncellenir. (n) mini-yığındaki eleman sayısı olmak üzere güncelleme işlemi denklem 2.6 ile hesaplanır.

$$\theta = \theta - \eta \cdot \nabla \theta J(\theta; X_{i+n}; y_{i+n}) \quad \textbf{2.6}$$

Gradyan iniş algoritması ve 3 farklı varyasyonunun en iyileme süreci şekil 2.5 de temsili olarak gösterilmiştir.

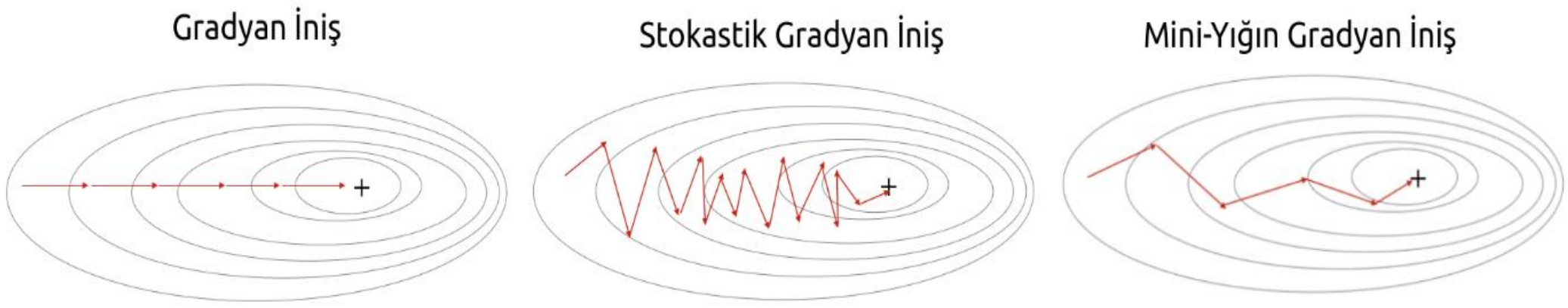

Şekil 2.5. Gradyan iniş yöntemlerinin temsili eniyileme karakteristiği. (Jacques M., 2020)

## 2.4 Kayıp Fonksiyonları

Kayıp fonksiyonları, tasarlanan modelin eğitimi sırasında problemin çözümüne yaklaşmaya teşvik etmek ve modelin başarısını değerlendirmek için kullanılır. Eğitim sırasında modelin eğitim kümesi üzerindeki veya test veri kümesindeki başarısını değerlendirerek eğitim sürecinin başarılı devam edip etmediği ve modelin genelleştirme yeteneği hakkında bilgi edinilmesine olanak tanır. Ayrıca öğrenme sürecinde modelin yaptığı hatanın hesaplanması, modelin kendi hatalarını düzeltmesi için en önemli adımlardan bir tanesidir. Bu bölümde görüntü verileri ile ilgili birkaç örnek kayıp fonksiyonundan söz edilecektir.

### 2.4.1 Manhattan mesafesi

L1 mesafesi Manhattan mesafesi olarak da bilinmektedir. Beklenen çıktı ile gerçek çıktı birbirinden çıkarılır ve sonucun mutlak değeri alınarak toplanır. L1 mesafesi denklem 2.7'de gösterilmektedir.

$$L\ (X_1, X_2) = \|X_1 - X_2\| \qquad \textbf{2.7}$$

### 2.4.2 Öklid mesafesi

L2 mesafesi öklid uzaklığı olarak da bilinir. Beklenen çıktı ile gerçek çıktının farklarının karesi alınıp toplanır ve sonuca karekök işlemi uygulanır. Bu kayıp fonksiyonunun girdi görüntülerini yeniden elde etme (reconstruction) görevi sırasında bulanık çıktılara neden olur (Pathak et al., 2016; Zhang et al., 2016). Öklid mesafesi denklem (2.8) ile ifade edilebilir.

$$L\ (X_1, X_2) = \ \|X_1 - X_2\|_2 \qquad \textbf{2.8}$$

### 2.4.3 Cosine benzerliği

Cosine benzerliği iki vektörün arasındaki açıyı ölçerek vektörler arasındaki benzerliği veya mesafeyi ölçmeyi amaçlar. Denklem 2.9'da gösterilmiştir.

$$L\ (X_1, X_2) = \frac{X_1 . X_2}{(\|X_1\|^2 * \|X_1\|^2} \qquad \textbf{2.9}$$

### 2.4.4 Algısal kayıp

Algısal kayıp fonksiyonu genellikle iki farklı girdininin VGG (Simonyan and Zisserman, 2014) ağının ara katman çıktıları arasında ağırlıklı kayıpların toplamı şeklinde hesaplanır. Bu yöntem genellikle görseldeki yerel özelliklerin korunabilmesi için kullanılır. Algısal kayıp fonksiyonu denklem 2.10 ile ifade edilebilir.

$$L\ (X_1, X_2) = \ L_1(Vgg(\ X_1), Vgg(\ X_2)) \qquad \textbf{2.10}$$

### 2.4.5 Kimlik kaybı

Kimlik kayıp fonksiyonu görseldeki kimlik bilgisi üzerinden bir kayıp değeri hesaplama durumlarında kullanılır. Örneğin üretilen görüntüdeki kimlik bilgisi ile girdideki kimlik bilgisinin aynı olmasının istediği durum için örnek bir kimlik kayıp fonksiyonu, (ID) kimlik bilgisi üreten bir model olmak üzere denklem 2.11 ile gösterilebilir.

$$L\ (X_1, X_2) = \ L_1(ID(\ X_1), ID(\ X_2)) \qquad \textbf{2.11}$$

# 3. ÜRETİCİ MODELLEME

Bu bölümde, üretici modellemenin temel amacı ve kullanılan yöntemlere kısaca değinilmektedir. Tez ile ilgili olarak çekişmeli üretici ağlar ile üretici modelleme yöntemleri detaylıca ele alınmaktadır.

Makine öğrenmesinde kullanılan en önemli paradigmalardan birisi tanımlayıcı modellemedir. Tanımlayıcı modellemeye örnek olarak sınıflandırma problemi verilebilir. Sınıflandırma probleminde bir örnek girdi olarak alınır ve model onun hangi sınıfa ait olduğunu tahmin etmeye çalışır. Örneklerden sınıfların öğrenildiği bu süreç bir tanımlayıcı modelleme örneğidir. Üretici modellemede ise bu örneğin tam tersi bir durum söz konusudur. Sınıfı girdi olarak verilen sınıfa ait bir sentetik örnek üretilmeye çalışılır. Üretici modellemede girdi örneklerinin hangi dağılımdan olduğu değil, dağılımın kendisinin öğrenilmesi amaçlanır.

Üretici modelleme için Enerji tabanlı yöntemler (LeCun, Y. et al., 2006;Du, Y. et al., 2019; Ho, J. et al., 2020), Varyasyonel otomatik kodlayıcılar (VAE) (Child, R., 2020; Kongma, D. P. et al., 2013 ), Çekişmeli üretici ağlar (GANs) (Goodfellow, I. et al., 2014; Radford, A. et al., 2015; Karras, T. et al., 2017, Karras, T. et al., 2021 ) ve daha birçok yöntem (Fahlman et al., 1983; Ackley et al., 1985; Hinton et al., 1984; Hinton ve Sejnowski, 1986; Hinton et al., 2006; Kingma et al., 2013; Rezende et al; 2014;) söz konusudur. Bu bölümde, son zamanlardaki en başarılı yöntemlerden kabul edilen ve bu tez kapsamında yapılan çalışmanın temelini oluşturan çekişmeli üretici ağlardan bahsedilecektir.

## 3.1 Çekişmeli Üretici Ağlar

Çekişmeli Üretici Ağlar (GANs), 2014 yılında I. Goodfellow tarafından tanıtılmıştır (Goodfellow et al., 2014). Tanıtıldıktan bugüne araştırmacıların ilgisini üzerine çekmeyi başarmıştır. Böylece kendisine bilgisayarlı görü (Ma, L. et al., 2017;Yu, J. et al., 2018) ve doğal dil işleme (Patashnik, O. et al., 2021; Zhang, H. et al., 2017, Yang, Z. et al., 2017) gibi birçok alanda kullanım alanı bulmuştur.

GAN'lar diğer üretici modeller ile karşılaştırıldığında dağılımları öğrenebilme, istenilen örnekleri verimli bir şekilde üretebilme gibi bir çok yönden avantajlara sahiptir. Bu avantajları özellikle bilgisayarlı görü alanındaki süper çözünürlük (super resolution), görselden görsele geçiş (image-to-image translation) ve görüntü tamamlama (image completion) gibi önemli problemlerde yüksek başarı elde edebilmesinde yardımcı olmuştur (Wang, X. et al., 2021; Wang, T. C. et al., 2018;Yeh, R. A. et al., 2017). Ancak GAN'lar da kendi problemlerine sahiptir. En önemli iki probleminden bir tanesi zor eğitilmesi diğer problem ise değerlendirmesinin zor olmasıdır. (Wang, Z. et al., 2020).

Çekişmeli üretici ağlarda, temelde iki farklı ağ birbiriyle mücadele ederken, gerçekçi sentetik veri üretmeyi öğrenirler. Bu ağlardan bir tanesi üretici (generator) olarak adlandırılır ve gerçekci sentetik veriyi üretir. Diğer ağ ise ayırıcı (discriminator) olarak adlandırlır. Bu ağın görevi üretici ağın ürettiği çıktıların gerçekçiliğini denetlemektir. Verimli bir öğrenme süreci gerçekleştirebilmek için iki ağ arasında bir denge yakalanması gerekmektedir. Dengeli bir eğitim geçirilmeyen durumlarda mod çökmesi (mode collapse) adı verilen bir problem ile sıkça karşılaşılır. Bu problem üretilen sentetik verilerin genel dağılımın sadece dar bir alt kümesinin üretilebildiği yani üretilen örneklerdeki varyansın düşük olduğu duruma verilen addır. Üretici modellemede yüksek çeşitlilikte sentetik verilerin üretilmesi hala üzerinde çalışılan önemli araştırma konusudur (Pan, Z. et al., 2019).

G, üretici, D ayırıcı z normal dağılımdan örneklenmiş bir vektör olmak üzere çekişmeli üretici ağların en genel amaç fonksiyonu denklem 3.1'de gösterilmiştir.

$$min_G max_G V(D,G) = E_x \sim p_{data}(X)\ [log\ D(X)] + E_z \sim p_Z(z)\big[log\ (1 - D\big(G(z)\big))\big] \tag{3.1}$$

Öğrenme süreci çekişmeli (adversarial) olacak şekilde ilerler. Bu çekişme süreci bir sıfır toplamlı oyun şeklindedir. Üretici ağ başarılı sonuçlar ürettiğinde ayırıcı ağ gerçek ve sahteyi ayıramaz ve maliyet fonksiyonun değeri yükselir. Bu durumda ayırıcı ağ, ayırma işlemini daha iyi öğrenebilmesi yönünde teşvik edilmiş olur. Ayırıcı ağ sentetik ve gerçek verileri yeterli doğrulukta ayırt edebildiğinde üretici ağ, ayırıcı ağı yanıltamaz. Bu durumda da üretici ağın maliyeti yükselir. Bu durumda da üretici ağ daha gerçekçi çıktılar üretebilmesi konusunda teşvik edilmiş olur. Böylece iki ağ, çekişmeli bir şekilde birbirlerini eğiterek gerçekçi sentetik çıktıyı üretmeyi öğrenirler. Üretici ve Ayırıcı ağ arasındaki bu ilişki, GAN eğitimini zorlaştırmaktadır. İki ağın kapasitelerinin dengeli olmadığı, öğrenme hızlarının farklı olduğu ya da en iyilemeye uygun bir noktadan başlamadıkları gibi nedenlerle arasındaki denge bozulabilmektedir. Bu durumda stabil bir eğitim sürecini elde edilemediğinden optimal kalitede çıktılara ulaşılamamaktadır.

GANs tanıtıldığı günden bu yanada üzerinde sıkça çalışılmış ve birçok versiyonu geliştirilmiştir. Genel olarak çalışmalar daha stabil ve başarılı eğitim süreçleri elde etmek (Arjovski et al., 2017; Gulrajanı et al., 2017; Kerras et al., 2017) ve ilgili göreve uygun özel mimariler (Kerras et al., 2019; Gecer et al., 2019; Choi et al., 2020, Kupyn et al.2018, Wu et al., 2019) tasarlamak şeklinde iki ana kategoriye ayrılabilir (Gui et al., 2020). Şekil 3.1'de standart bir GAN yapısı gösterilmektedir.

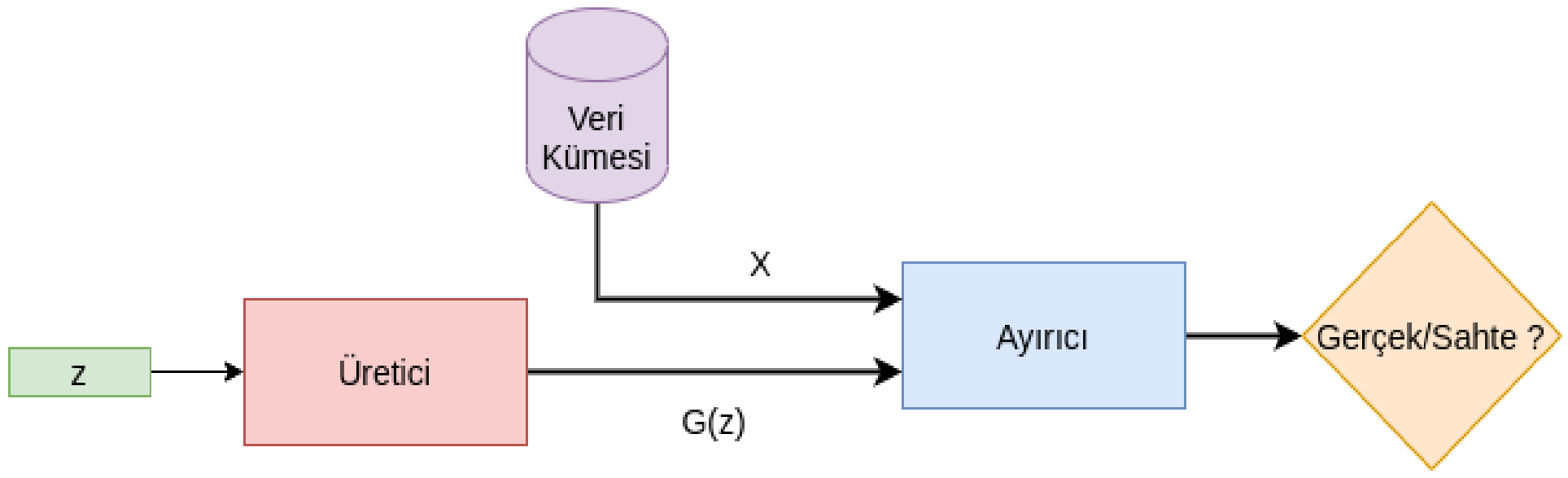

Şekil 3.1. Standart GAN yapısı

## 3.2 Koşullu Çekişmeli Üretici Ağlar

Koşullu üretici ağlar (Mirza et al., 2014) genellikle üretim sürecini kontrol edebilmek için kullanılır. Üretici ağ'ın ilgili girdisine ek olarak üretilen çıktıların kontrol edebilecek ek girdiler sağlanarak üretici ağın üretme süreci bu girdiğe bağımlı hale getirilir. Böylelikle kontrol girdisi veya girdileri ile birlikte istenilen çıktı elde edilebilir. Koşullu çekişmeli ağlar için genel amaç fonksiyonu denklem 3.2'de verilmiştir.

$$min_G max_G V(D,G) = E_x \sim p_{data}(X)\ [log\ D(X|y)] + E_z \sim p_Z(z)\left[\log\left(1 - D\big(G(z|y)\big)\right)\right] \quad \textbf{\textit{3.2}}$$

Koşullu çekişmeli ağların temel yapısı Şekil 3.2'de gösterilmektedir.

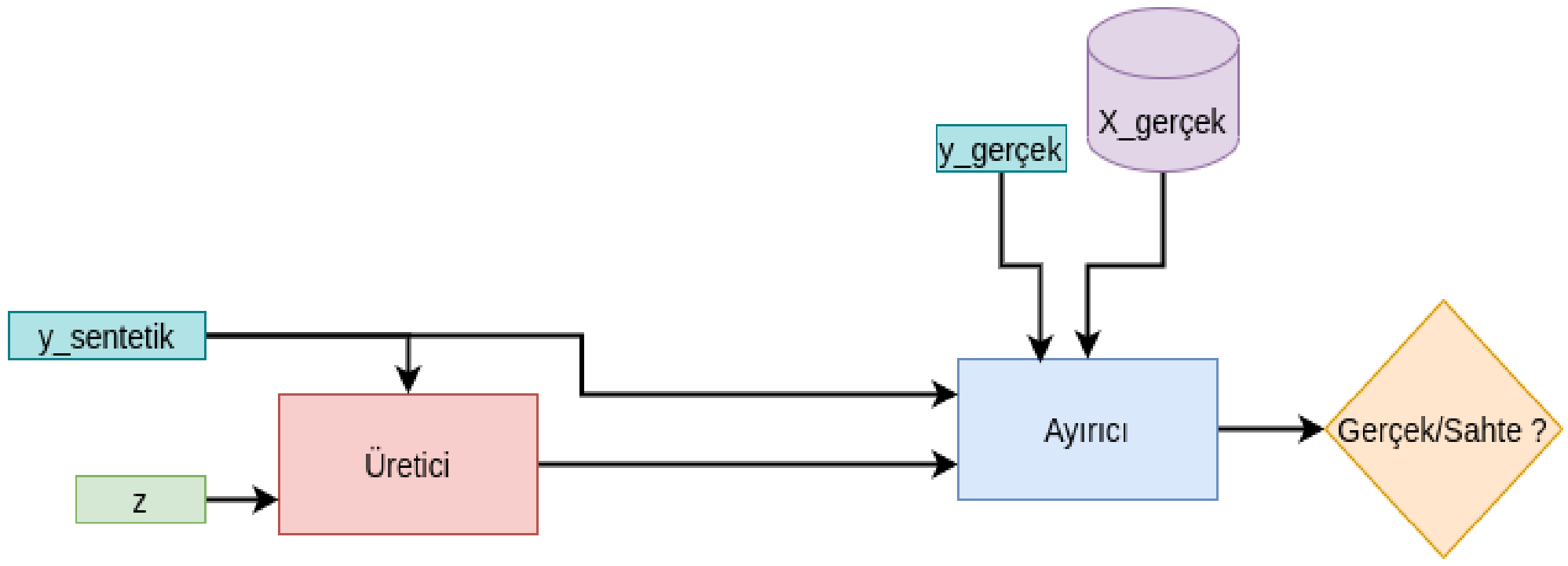

Şekil 3.2. Koşullu GAN temel yapısı

# 4. ÖNCEKİ ÇALIŞMALAR

Bu bölümde çekişmeli üretici ağlar ile saç sentezi literatürü incelenecektir. İlgili çalışmalar, problemin çözümü için kullanılan veri kümeleri ve sonuçların değerlendirilme yöntemleri üzerinde durulacaktır.

## 4.1 İlgili Çalışmalar

Saç sentezi problemi farklı amaçlar ile birçok farklı yönden ele alınabilir. Yüz görüntülerinde görüntü düzenleme (image editing) problemi üzerinde yapılan bazı çalışmalarda görüntünün farklı yerleri çeşitli şekillerde değiştirilir. Örneğin saç ve yüzün belli bölgeleri eskizler ile değiştirilerek modele girdi olarak verilir. Elde edilen çıktıda gerçekçi bölgeler korunurken eskiz ile değiştirilen bölgeler için eskize uygun şekilde gerçekçi çıktılar üretilir. Daha sonra önerilen çözüm ile değiştirilen bölgeye uygun gerçekçi çıktı tahmin edilmeye çalışır. Bu ve benzeri şekillerde yüz üzerinde duran saçlar uygun bir şekilde kaldırılabilir yada saç bölgesi farklı saç şekilleri ile değiştirilebilir. (Yang et al., 2020; Portenier et al., 2018, Jo et al., 2019)

Saç sentezleme problemi üzerinde hali hazırda, içinde saç'a ait bölgeler de barındıran gerçekçi çıktılar üreten bir çekişmeli üretici ağın, vektör uzayında düzenlemeler yapılarak da çalışabilir. Zhu ve arkadaşlarının Barbershop adlı çalışması ile Saha ve arkadaşlarının LOHO adlı çalışmasında, vektör uzayında yapılan iteratif optimizasyon yöntemleri ile saç stil transferi ve saç sentezi problemi üzerinde çalışılmıştır. (Zhu et al., 2021; Saha et al., 2021).

Koşullu üretici ağlar ile farklı koşullar için uygun, kontrollü bir şekilde saç üretebilmek üzerine yapılan çalışmalar da bulunmaktadır. Tan ve arkadaşları, sözde denetimli öğrenme (pseudo supervised learning) stratejisini kullanarak, geliştirdikleri koşullu çekişmeli üretici ağ yöntemi ile saç sentezleme işleminde şekil, yapı, görünüm ve arkaplan özelliklerinin birbirinden bağımsız bir şekilde kontrol edilebilmesi üzerinde çalışmışlardır. (Tan et al., 2020). Fan ve arkadaşlarının yaptığı çalışmada geliştirilen koşullu üretici ağ yönteminde ise saçın rengi, yapısı ve şekli kontrol edilebilmektedir (Fan et al., 2021).

Çekişmeli ağlar ile saç üretme problemi Olszewski ve arkadaşlarının çalışmalarında iki faz olarak ele alınmıştır. İlk fazda sentetik verilerden oluşan bir veri kümesi kullanarak saçın üretilmesi ve ikinci fazda ise gerçek saç görüntülerinden oluşmuş bir veri kümesi ile çıktıların iyileştirilerek gerçek görüntüye yerleştirilmesi şeklinde ele alınmıştır. (Olszewski et al., 2020).

Metinler kullanılarak belirtilen ifadele uygun saçlı görseller üretmek için hem metin hem görsel verilerin kullanıldığı çalışmalar da mevcuttur. Bu çalışmalarda amaç yazılan metne uygun renk ve görünüme sahip saç görsellerin üretilmesidir (Wei et al., 2021).

## 4.2 Veri Kümeleri

Çekişmeli üretici ağlar ile saç sentezi literatürü oldukça yeni ve dar bir alandır. Saçın dinamik yapısı, dokusu gibi kendine has özelliklerinin yanı sıra bir kişinin farklı saç stillerinde bulunmasının zor bulunan bir durum olması, bu alana özel bir veri kümesi oluşturulmasının önünde önemli bir engeldir. Bildiğimiz kadarıyla bu problem için hazırlanmış, kullanıma açık bir veri kümesi bulunmamaktadır. Bu nedenle saç sentezi problemi çözümü için araştırmacılar farklı yollar izlemektedirler.

Bazı araştırmacılar online uygulamaları (Laticis Imagery., 2017; Daz Productions., 2017) kullanarak sentetik veri kümesi oluşturmuş ve bu sentetik veri kümesinde aldığı sonuçları daha küçük ama gerçek verilerden oluşmuş bir veri kümesi ile iyileştirmeyi hedeflemişlerdir. (Olszewski et al., 2020) Bazıları ise yüksek kaliteli az sayıda görüntü üzerinde çalışmalarını gerçekleştirmişleridir. (Zhu et al., 2021)

Bazı araştırmacılar (Tan et al., 2020; Jo et al., 2019) ise yüz'ü odağına alan problemlerinin çözümü için kullanılan FFHQ (Kerras et al., 2019), CelebA-HQ (Kerras et al., 2018) gibi bazı iyi bilinen veri kümelerinden faydalanmışlardır.

## 4.3 Değerlendirme Yöntemleri

Bu bölümde, saç sentezi probleminin çözümü için sunulan yöntemlerin başarılarının karşılaştırılması ve kullanılan yöntemlerin sonuçlarının değerlendirilmesi için kullanılan nitel ve nicel yöntemler üzerinde durulacaktır.

Nicel yöntemler, matematiksel bir metrik ile sonuçların değerlendirildiği yöntemlerdir. Bu yöntemler ile açıkça sonuçların başarısı değerlendirilebilir. Bu sayede çıktı olarak elde edilen nicel değerleri kullanarak hassas bir şekilde farklı çözümleri karşılaştırmak mümkün hale gelir. Saç sentezleme probleminin çözümünde elde edilen veri, görüntü olduğu için genellikle bu problemin başarısında görüntü değerlendirme metrikleri kullanılır. Bu metrikler, doğrudan tahmin edilen görüntüyle gerçek görüntünün arasındaki farkı hesaplayabildiği gibi bazıları da dağılımlarını karşılaştırabilir. Bu metriklere RMSE, PSNR, SSIM (Wang, Z. et al., 2004), LPIPS(Zhang et al., 2018), FID (Heusel et al., 2017), L1, L2 örnek olarak verilebilir.

Gerçekçi saç sentezi problemi, görsel çıktılar üretilen bir problem olduğu için nitel değerlendirme de çoğunlukla kullanılır. Bu değerlendirme şeklinde önerilen yöntem ve diğer yöntemlerin aynı girdiler için aynı amaca yönelik çıktıları istenir ve elde edilen çıktılar kişisel gözlemler ile değerlendirilir. Bu gözlemler özellikle nicel olarak yeterince değerlendirilemeyen problemlerin tespitinde sıklıkla kullanılır. Örneğin çıktılardaki yapay hatalar (artifact) gözlenir, ışık gibi görsel durumlar incelenebilir.

Bu bölümde belirtilen doğrudan saç sentezi problemine odaklanan yöntemler, kullandıkları temel yaklaşımlar ve sonuçlarını değerlendirme yöntemleri Tablo 4.1'de sunulmuştur.

Tablo 4.1. Literatür araştırması özeti

| Çalışma | Yöntem | Veri Kümesi | Değerlendirme Yöntemi |
|---|---|---|---|
| (Zhu et al., 2021) | StyleGANv2 vektör uzayını görsel özelliklerin daha verimli taşınabildiği bir vektör uzayına dönüştürülerek bu uzayda saç stili probleminin çözümü üzerinde çalışılmıştır. Çalışma zamanında iteratif çalışan bir yöntemdir. | 120 adet 1024x1024 çözünürlükte görseller kullanılmıştır. (Zhu et al., 2020) | Nitel gözlemler ve PSNR, RMSE, SSIM, VGG, LPIPS, FID gibi nicel metrikler. |
| (Saha et al., 2021) | Saç stil transferi yaparken aynı zamanda vektör uzayında saç'a ait detayları doldurabilir. Saçın algısal yapısı, görünümü ve stil bilgisini bir birinden ayırır ve her bir özellik için belirlenen kayıp fonksiyonları ile çalışma zamanında optimizasyon işlemi yapılır. | 70.000 adet yüksek kaliteli görüntüden oluşan FFHQ veri kümesi kullanılmıştır. | FID, PSNR, SSIM gibi nicel metriklerin yanında görsel değerlendirmelerde bulunulmuştur. |
| (Tan et al., 2020) | Saçın görünümü, arkaplanı, şekli ve yapısının bir birbirini etkilemeden değişebildiği bir koşullu çekişmeli üretici ağ yöntemidir. Ayrıca düzemleme yapabilemek için bir arayüz uygulaması da geliştirimiştir. | 70.000 adet yüksek kaliteli görüntüden oluşan FFHQ veri kümesi kullanılmıştır. | Nitel değerlendirmelerin yanı sıra nicel bir metrik olarak FID'den faydalanılmıştır. |
| (Wei et al., 2021) | Verilen referans görsel veya metin kullanılarak orjinal görüntüdeki saç bölgesinin düzenlenebildiği bir yöntem sunulmaktadır. | 30.000 adet yüksek kaliteli görüntüden oluşan Celeba-HQ very kümesi kullanılmıştır. | IDS, PSNR ve FID gibi metriklerle birlikte nitel değerlendirmeler de kullanılmıştır. |
| (Fan et al., 2021) | Saçın renk, yapı ve şeklinde düzenlemeler yapmaya izin veren bir koşullu çekişmeli üretici ağ yöntemidir. | 70.000 adet yüksek kaliteli görüntüden oluşan FFHQ veri kümesi kullanılmıştır. | IDS, PSNR, ACD ve SSIM gibi nicel metriklerin yanında nitel değerlendirmeler de yapılmıştır. |

# 5. ÇEKİŞMELİ ÜRETİCİ AĞLAR İLE SAÇ SENTEZİ

Bu bölümde saç sentezi için önerilen, çekişmeli üretici ağlar temelli yöntemimizin genel yapısı, alt modülleri ve eğitimi için kullanılan kayıp fonksiyonları hakkında bilgi verilecektir.

Bu çalışma kapsamında önerilen yöntemin gerçeklenmesi için Python programlama dili ve yapay zeka araştırmalarında sıkça kullanılan Pytorch kullanılmıştır.

## 5.1 Genel Bakış

Günümüzde görüntülerde gerçekçi saç sentezleme işlemi önemli alanlarda kullanılabilmektedir. Bu işlem, saç rengi ve şeklini değiştirerek kimliği gizlenen kayıp veya suçlu bireylerin olası durumlarının tahmininde yada gelecekte görsel verilerde anonimliğin sağlanması için yapılan veri işlemlerinde destekleyici etken olarak kullanılabilecektir. Ayrıca kişisel görünümünde değişiklik yapmak isteyen kişiler, uzun süreli fiziksel değişime maruz kalmadan olası durumlarını görüp, önceden bir değerlendirme yapabilirler. Tüm bu ve benzeri kullanım alanlarının yanı sıra gerçekçi saç sentezleme problemi, saçın doğası itibariyle zorlu bir problemdir. Bu problem içerisinde bir çok zorlu alt problem içerir. İnsan yüzünün diğer bölümlerinin aksine kendine özgü geometrisi ve materyali saçın analiz, temsil ve üretimini zorlaştırmaktadır. Daha detaylı incelendiğinde zorlu saç maskesi sınırları, hassas saç şekli kontrolü ihtiyacı, arka plan ile saçın kusursuz kesişimi, bu problemin temel zorluklarından bazıları olarak gösterilebilir. Çekişmeli üretici ağlar ile saç sentezi probleminin görece yeni bir alan olmasından dolayı, oldukça dar bir literatüre sahiptir. Biz de, bu tez kapsamında çekişmeci ağlar ile gerçekçi saç sentezinde literatürdeki en iyi yöntemler ile yarışır sonuçlar veren bir yöntem sunuyoruz.

## 5.2 Yöntem

Önerilen yöntem bir çekişmeli üretici ağ mimarisidir (Şekil 5.1). Yöntem, bir kodlayıcı ağ, bir üretici ağ ve bir ayırıcı ağdan oluşmaktadır. Üretici ağ iki kısımdan oluşur. İlk kısım saç üretim katmanlarından ikinci kısım ile saç karıştırma katmanından oluşmaktadır. Üretici ağın ilk kısmı, gerçekçi saçların üretilmesinden sorumludur ikinci kısmı ise üretilen saç'ın arka plan ile kusursuz bir şekilde karıştırılması ve son düzenlemelerinin yapılması görevini üstlenir.

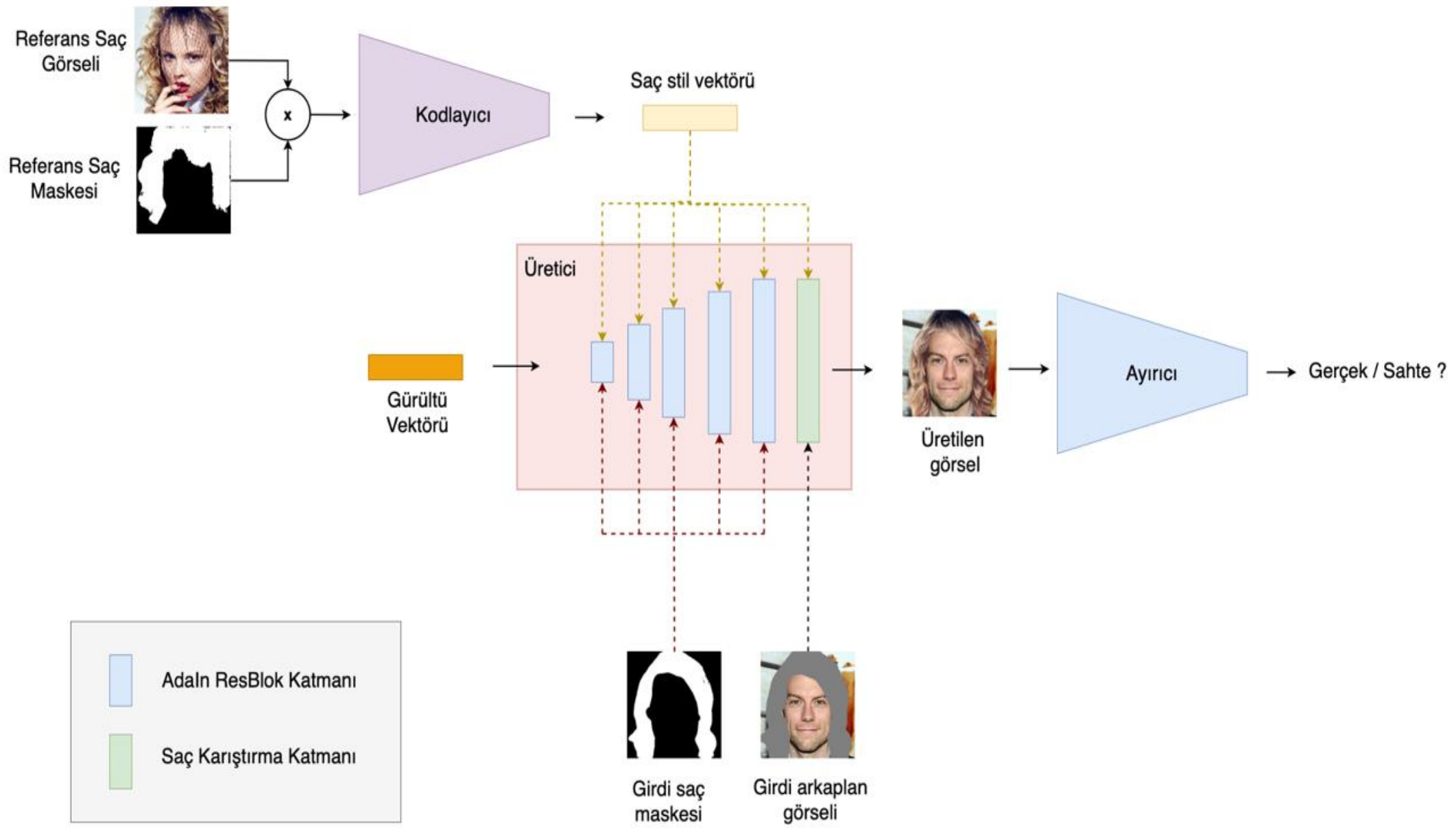

Şekil 5.1. Geliştirilen yönteme genel bakış

Kodlayıcı ağ, çekişmeli üretici ağ'ın koşul girdisini üretmektedir. Bu girdi üretilmek istenilen saçı temsil eden referans saçtan elde edilmiş bir vektördür. Referans saç'a sahip görsel, saç maskesi ile maskelendikten sonra referans saç bölgesi elde edilir. Referans saç bölgesini girdi olarak alan kodlayıcı ağ, çıktı olarak saç bölgesinin vektör uzayda bir temsilini üretir. Bu temsil ve rasgele örneklenmiş gürültü vektörü üretici ağ'ın girdisidir. Ayırıcı ağ ise üretici ağ'ın çıktılarının gerçekliğini değerlendirmekle yükümlüdür. Geliştirilen yöntem şekil 5.1 de gösterilmiştir.

### 5.2.1 Saç üretici ağ

Üretici ağ (G), iki kısımdan oluşmaktadır. İlk kısım, girdi olarak aldığı gürültü vektörü ile referans saç'tan elde edilen stil vektörünü kullanarak saç sentezi işlemini gerçekleştirmektedir. İkinci kısım ise, sentezlenen saç ve arkaplan görüntüsünün karıştırılmasından ve son düzenlemelerinin yapılmasından sorumludur. Gürültü vektörü, normal dağılımdan örneklenir ve $p_z(x)$ şeklinde ifade edilebilir. Stil vektörü ise s ile ifade edilebilir. Sonuç olarak gürültü vektörü ve kontrol değişkenini gerçekçi saç sentezine dönüştüren üretici ağ, denklem 5.1 ile ifade edilebilir.

$$X = G(z|s) \qquad \textbf{5.1}$$

Üretici ağın ilk kısmı, AdaInResBlok katmanlarından oluşmaktadır.

### 5.2.1.1 AdaINResBlok katmanı

Adaptif Örnek Normalizasyonu (AdaIN) (Kerras et al.,2019; Dumoulin et al., 2016; Dumoulin et al., 2018; Ghiasi et al., 2017) , bir önceki katmandan gelen içerik girdisini ve stil vektörünü girdi olarak alır ve içerik girdisini istatistiksel olarak stil vektörüne hizalar. Hizalama işlemi basitce ortalama( μ) ve standart sapma(σ) değeri ile normalize edilmiş içerik girdisinin, stil vektörünün standart sapmasıyla ölçeklenerek, yine kontrol değişkeninin ortalaması kadar ötelenmesidir. AdaIN işlemi denklem 5.2 ile ifade edilebilir.

$$AdaIN\ (X, y) = \sigma(y)\left(\frac{X - \mu(X)}{\sigma(X)}\right) + \mu(y) \qquad \textbf{5.2}$$

AdaInResBlok katmanı (Choi et al., 2020), RestBlok katmanına AdaIN normalizasyonunun entegre edilmiş versiyonudur. Bu katmanda kontrol değişkeni, AdaIN vasıtasıyla ağ’a enjekte edilir. Böylelikle üretici ağın katmanlarında kontrol değişkenine istatistiksel olarak bir yakınlaşma söz konusu olur. AdaINRestBlok katmanı, girdisini iki kez, adaptif örnek normalleştirmesi, aktivasyon fonksiyonu ve evrişim katmanı üçlüsünden geçirir. Elde edilen çıktı, girdinin başka bir evirişim katmanı çıktısı ile toplanarak sonuca aktarır. Şekil 5.2’de yöntemimizde kullanılan AdaINResBlok gösterilmektedir.

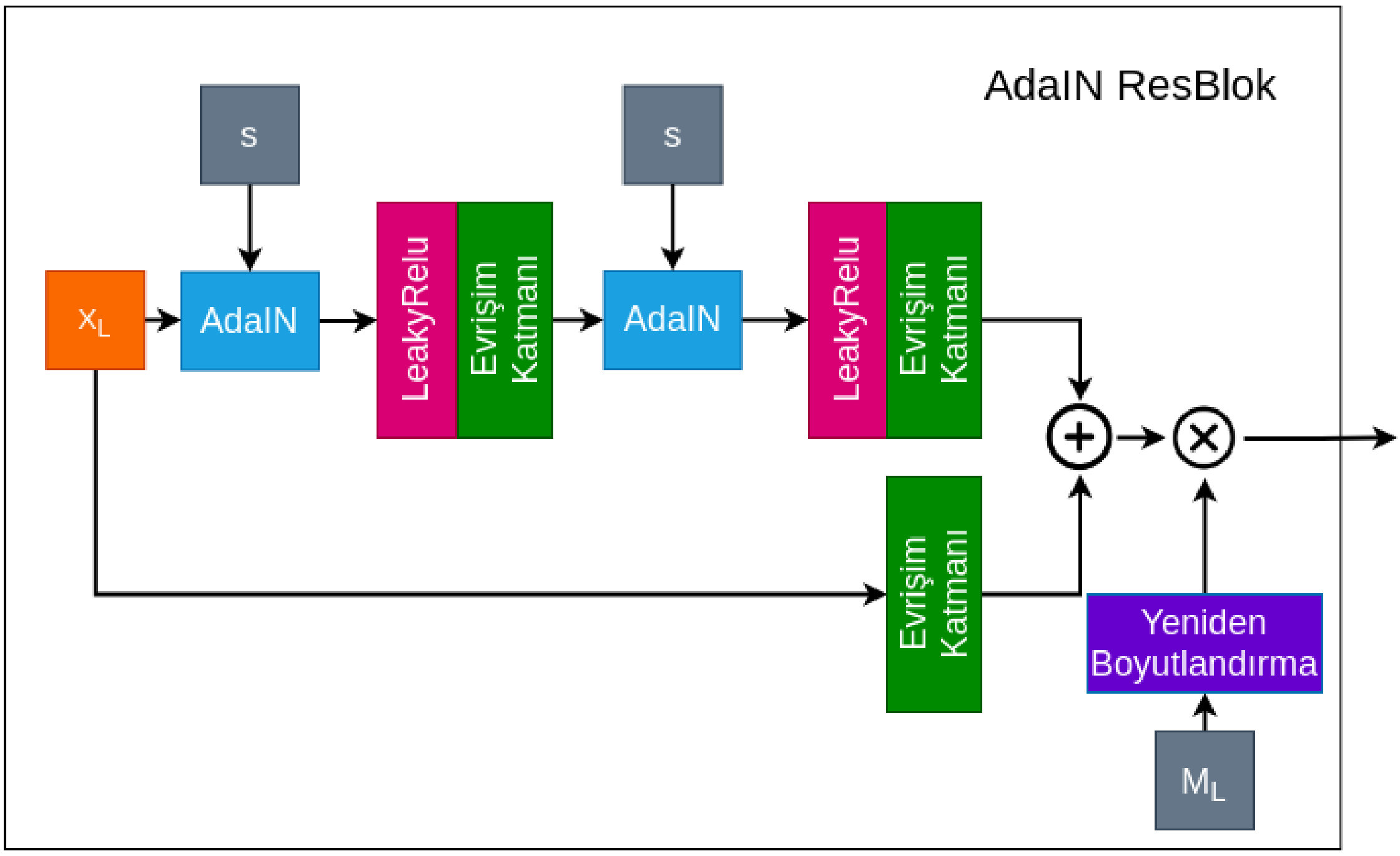

Şekil 5.2. AdaINResBlok katmanı

AdaInResBlok'lardan oluşan üretici ağ'ın ikinci kısmı saç karıştırma katmanıdır. Bu iki kısım birlikte üretici ağ'ı oluşturmaktadır.

### 5.2.1.2 <u>Saç karıştırma katmanı</u>

Gerçekçi saç sentezleme probleminin en önemli alt problemlerinden birisi üretilen saç ile arkaplan görüntüsünün kusursuz bir şekilde karıştırılmasının sağlanmasıdır. Bu problemin çözümü için, alfa ve poisson karıştırma yöntemi gibi yöntemler kullanılabileceği gibi üretici ağ'ın tamamı veya ek bir ağ ile de karıştırma problemini çözümü üzerinde çalışmak mümkündür (Pérez, P. et al., 2003; Zhang, L. et al. 2020; Wu, H. et al., 2019). Bu tez kapsamında önerilen yöntem, gerçekçi saç sentezi yapılırken aynı zamanda verimli bir yöntem olmasıda beklendiğinden önerilen yöntemde ek bir ağ kullanılması tercih edilmemiştir. Onun yerine üretici ağ'ın son katmanını oluşturan saç karıştırma katmanı öneriyoruz. Bu katmanda içerik girdisi AdaIN ResBlok'tan geçirdikten sonra saç maskesi ile çarpılır ve orjinal görüntüdeki arkaplan bölgesiyle toplanır. Elde edilen çıktı bir evrişim katmanından geçirilerek üretici ağ'ın nihai çıktısı elde edilmiş olur. Bu karıştırma işleminin üretici ağ'ın sonunda gerçekleştirilmesi, ayırıcı ağ'ın karıştırılacak alanların gerçekçi bir şekilde düzenlemesini denetleyebilmesini sağlar. Saç karıştırma katmanı Şekil 5.3’de gösterilmektedir.

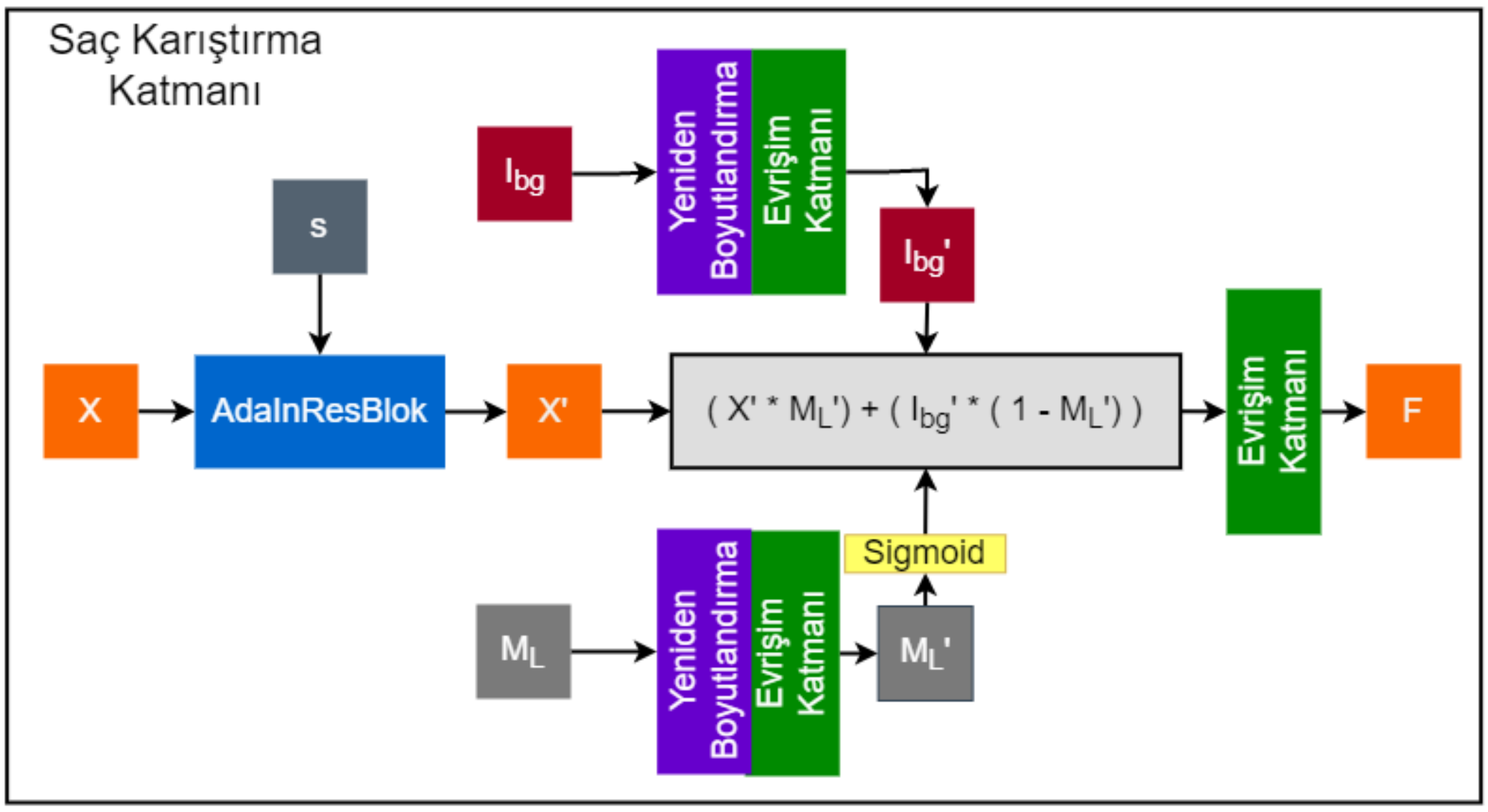

Şekil 5.3. Saç karıştırma katmanı

### 5.2.2 Ayırıcı ağ

L2 ve L1 kayıp fonksiyonları tek başına gerçekçi çıktılar üretmekte yeterli olamamakta, fotoğraf üretme probleminde bulanık çıktılara sebep olmaktadır (Larsen, A. et al., 2016). Bu kayıp fonksiyonları, yinede düşük frekans özellikleri yakalamada başarılılardır. (Isola, P. et al., 2017). Bu çalışamada kullanılan ayırıcı ağ evrişim katmanlarından oluşan bir kodlayıcı ağıdır. Bu ağ üretici ağ tarafından sentezlenen görüntülerin gerçekliğini değerlendirmek ve değerlendirme sonuçlarıyla üretici ağı daha gerçekçi çıktılar üretmeye teşvik etmekle yükümlüdür. Aldığı görüntü girdisini tek kanallı NxN’lik bir çıktıya eşleyerek, yerel özniteliklerin istatistiksel özelliklerini yakalamayı amaçlamaktadır. (Li and Wand., 2016; Isola et al., 2017). Ayırıcı ağ’ın aldığı görüntünün gerçekliğini değerlendirmesi, denklem 5.3’de ifade edilmektedir.

$$p = D(X) \qquad \textbf{5.3}$$

### 5.2.3 Eniyilenen kayıp fonksiyonları

$X \in R^{H\,x\,W\,x\,C}$, $s \in R^{512}$, $z \in R^{512}$ olmak üzere, önerilen yöntem aşağıdaki anlatılan kayıp fonksiyonlarını optimize etmektedir.

#### 5.2.3.1 Algısal kayıp fonksiyonu

Saç bölgesine ait yüksek seviyeli özniteliklerin korunabilmesi için önceden ImageNet (Deng et al., 2009) üzerinde eğitilmiş bir VGG19 (Simonyan and Zisserman., 2014) modelini kullanarak, algısal kayıp (Johnson et al., 2016; Gatys et al., 2016) değerini hesaplıyoruz. Bu kayıp fonksiyonunda sentetik ve gerçek görüntünün VGG19 ağının ara katmanlarındaki aktivasyonlarının birbirinden uzaklığı hesaplanır. Böylece üretici ağı gerçek saç ile daha uyumlu çıktılar üretmesi için teşvik etmektedir. Algısal kayıp fonksiyonu denklem 5.4'te gösterilmektedir.

$$L_{algısal} = L_1(Vgg(X) - Vgg(G(X))) \qquad \textbf{5.4}$$

#### 5.2.3.2 Stil kayıp fonksiyonu

Verilen saç bölgesine ait özelliklerin, vektör uzayda başarılı bir şekilde temsil edilebilmesi, üretilen çıktıların istenilen seviyede elde edilebilmesi açısından kritik öneme sahiptir. Stil kayıp fonksiyonu, saç bölgesini vektör uzayda bir temsile dönüştüren kodlayıcı ağı en iyi temsili üretecek şekilde teşvik etmektedir. Stil kayıp değeri, gerçek saç bölgesinden elde edilen stil vektörü temsili ile üretilen saç bölgesinden elde edilen stil vektörü arasında L1 mesafesi hesaplanarak elde edilir. Stil kayıp fonksyionu denklem 5.5 ile gösterilebilir. (Huang et al., 2018; Zhu et al., 2017).

$$L_{stil} = L_1(E(X_{saç}), E\ (X'_{saç})) \qquad \textbf{5.5}$$

#### 5.2.3.3 Piksel kayıp fonksiyonu

Algısal kayıp fonksiyonuyla yüksek frekanslı bölgeler yakalanabilse de, çıktının doğru renkte ve doğru ışıkta olması gibi düşük frekanslı özelliklerin yakalanabilmesi için piksel kayıp fonksiyonu kullanılmaktadır. Bu kayıp fonksiyonunda girdi ve çıktı resimlerinin pixel tabanlı farklı alınarak hesaplanır ve üretici ağı düşük seviyeli öznitelikleri yakalayabilecek şekilde üretim yapmaya teşvik eder. Piksel kayıp fonksiyonu, denklem 5.6'da gösterilmektedir.

$$L_{piksel} = L_1(X, X') \tag{5.6}$$

### 5.2.3.4 Çekişmeli kayıp fonksiyonu

Çekişmeli kayıp fonksiyonu çekişmeli üretici ağlar tabanlı yöntemimizin stabil bir şekilde eğitilmesi amacıyla kullanılan kayıp fonksiyonudur. Gerçek ve üretilen görsellerin ayırıcı ağ çıktılarına göre üretici ve ayırıcı ağ arasındaki çekişmeli eğitimin gerçekleşmesini sağlar.

$$L_{çekişmeli} = E_x\,[log\,D(X)] + E_{z,s,M}\,[log\,(1 - D(G(z, s, M)))] \tag{5.7}$$

### 5.2.3.5 Genel amaç fonksiyonu

Algısal kayıp, Stil kayıp ve Piksel kayıp fonksiyonundan oluşan önerilen yöntem için kullanılan genel amaç fonksiyonu denklem 5.8 ile gösterilmektedir.

$$min_G max_D(\lambda_{piksel} * L_{piksel} + \lambda_{stil} * L_{stil} + \lambda_{algısal} * L_{algısal} + \lambda_{çekişmeli} * L_{çekişmeli}) \tag{5.8}$$

# 6. DENEYSEL ÇALIŞMALAR

Bu bölümde, bu tez kapsamında Çekişmeli üretici ağlar ile gerçekçi saç problemi için önerilen yöntemi ve onun farklı konfigürasyonlarını değerlendirebilmek için hazırlanan deney ortamlarından ve elde edilen deney sonuçlarından bahsedilecektir.

## 6.1 Veri Kümesinin Hazırlanması

Deneysel çalışmalarda FFHQ veri kümesi kullanılmıştır. FFHQ veri kümesi 70000 adet yüksek kaliteli (1024x1024) görselden oluşur. Celeb-HQ veri kümesinden yaş, etnik köken ve arka plan açısından çok daha fazla varyasyona sahip olmakla birlikte şapka, güneş gözlüğü gibi birçok aksesuar da içermektedir. Flickr web sitesinden çekilen görüntüler otomatik olarak hizalanmış ve kırpılmıştır (Kerras et al., 2019).

MichiGAN ile adil bir karşılaştırma yapabilmek adına, yöntemimizin eğitiminde MichiGAN önişleme adımları izlenerek eğitime hazırlanmış veri kümesi kullanılmıştır. Kullanılan ön işleme adımları sonucunda elde edilen veri kümesi 56000 eğitim ve 14000 test olacak şekilde iki farklı kümeden oluşmaktadır. Görseller 512x512 çözünürlüğe indirilmiş ve saç'a ait bölgelerin ikili maskeleri çıkarılmıştır. Önerilen yöntem, 128x128'lik çözünürlükteki girdiler için tasarlandığından MichiGAN ön işleme prosedürüne ek olarak 128x128'e yeniden boyutlandırma işlemi uygulanmıştır. Belirlenen önişlemeler sonucunda elde edilen görüntülerden rastgele seçilmiş örnek şekil 6.1 de gösterilmişir.

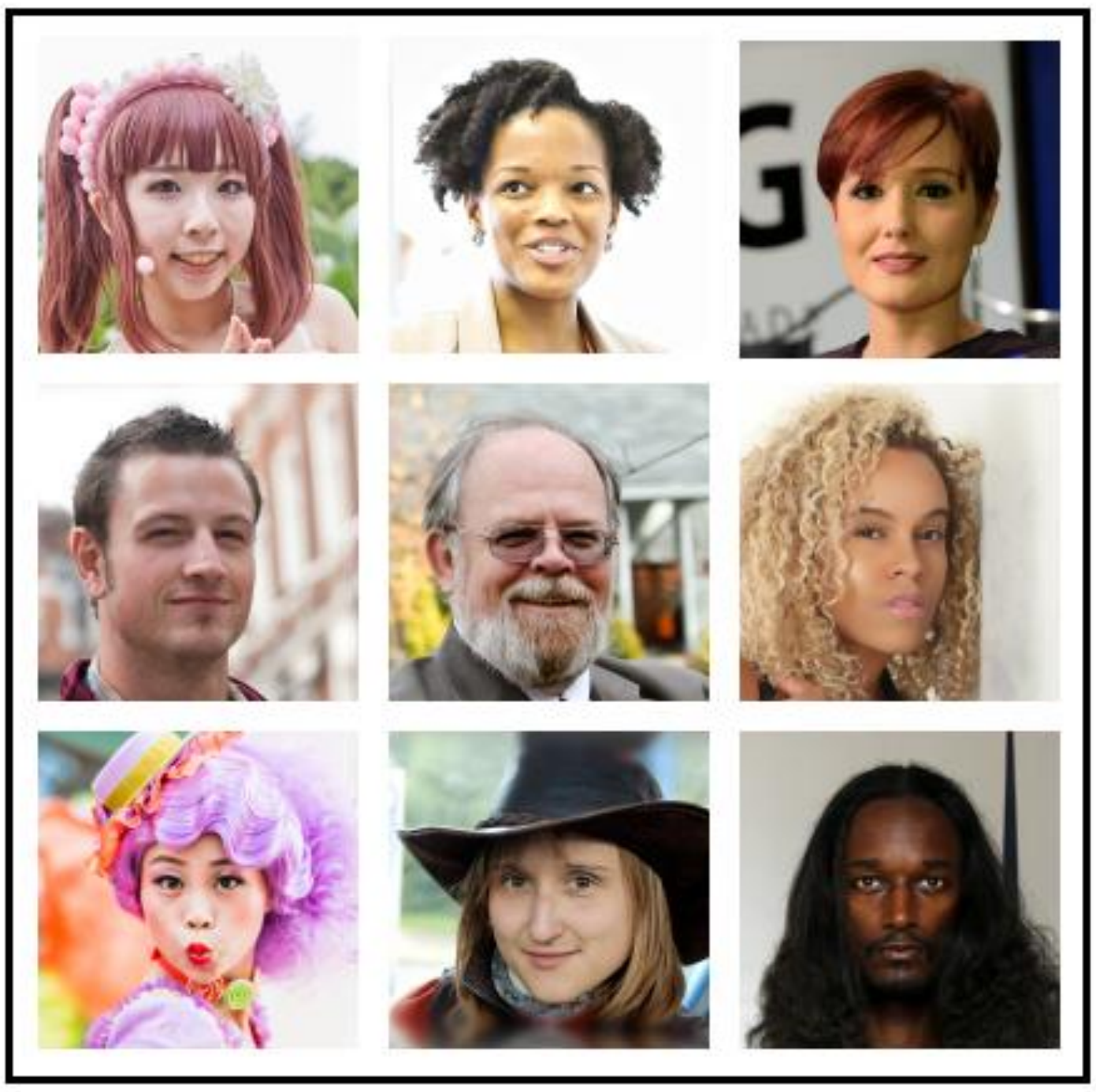

Şekil 6.1. FFHQ veri kümesinden rastgele seçilmiş örneklerin önişleme sonuçları.

Önerilen yöntemde saç bölgesini temsilen ikili saç maskesine ihtiyaç duyulmaktadır. Bahsedilen önişleme sonucunda saç olan bölgeler 1, olmayan bölgelerin 0 değerini aldığı ikili maskeler de elde edilmiştir. Bazı örnekler ve ikili maskeleri şekil 6.2'de gösterilmiştir.

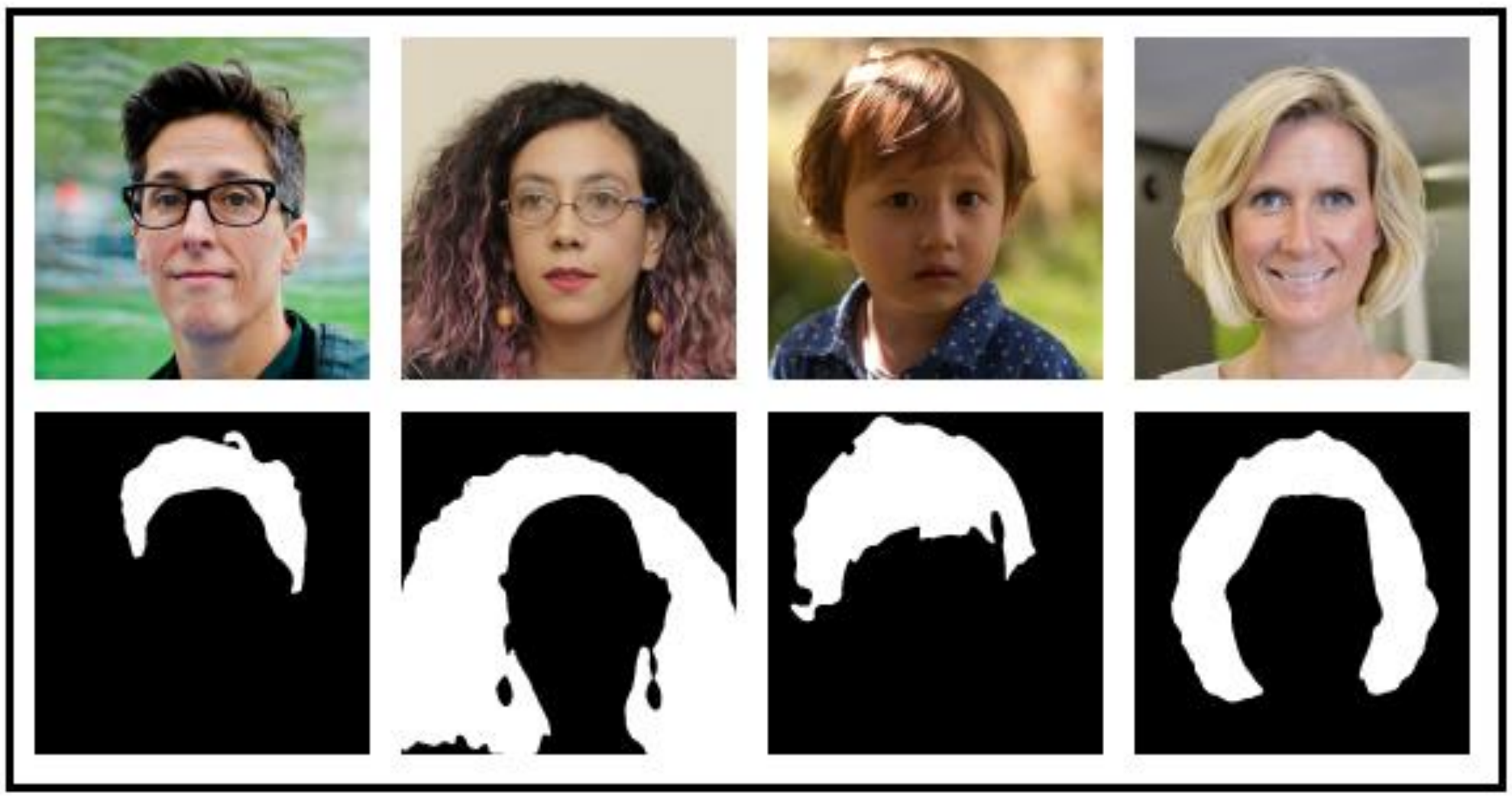

Şekil 6.2. FFHQ veri kümesinden rastgele seçilmiş bazı örnekler ve ikili saç maskeleri

## 6.2 Deney Ortamı

Tez kapsamında yapılan çalışmalarda 8GB belleğe sahip NVIDIA RTX2080 ekran kartı, intel core i9 9900K @ 3.6GHz 16 çekirdekli işlemci, 32 GB RAM ve Ubuntu 18.04 işletim sistemi ve CUDA 10.2 ye sahip bir bilgisayar kullanılmıştır.

Yöntemlerin gerçeklenmesi için Python programlama dilinin 3.6.7 sürümü ve Pytorch kütüphanesinin 1.8.1 sürümü kullanılmıştır.

## 6.3 Deney Sonuçları

Bu bölümde, önerilen yöntem ile Bölüm 6.1'de eğitim için hazırlanma süreci anlatılan veri kümesi kullanılarak deneyler yapılmıştır. Yapılan deneyler hem nicel hemde nitel olarak değerlendirilmiştir. Nitel ve nicel testlerin yapılması ve sonuçların değerlendirilmesi için bölüm 6.1 de belirtilen test kümelerindeki görseller kullanılmıştır. Önerilen yöntemin eğitimi sırasında, eniyileme yöntemi olarak Adam (Kingma et al., 2014) eniyileme algoritması kullanılmıştır. Üretici ağ ve ayırıcı ağ için öğrenme oranı 0.0001, B1 ve B2 parametreleri için sırasıyla 0.5 ve 0.999 değerleri kullanılmıştır. Eğitim süresi 55 dönem (epoch), yığın büyüklüğü (batch size) 8 olarak uygulanmıştır. Tüm kayıp fonksiyonu lambda değerleri 1 olarak belirlenmiştir.

### 6.3.1 Saç yeniden yapılandırma

Saç stili yeniden yapılandırma görevi, ilgili girdiler sağlandığınarak original saç görüntüsünü elde etmenin hedeflendiği bir görevdir. Saç stili yeniden yapılandırma görevinde girdi ve referans görüntü için aynı görsel seçilir. Önerilen yöntemin çalışması için gerekli girdiler saç maskesi, arka planı, saç bölgesi hazırlanır. Bu girdiler sonucunda orjinal görüntü model tarafından oluşturulmaya çalışılır. Bu işlem denklem 6.1 ile formülize edilebilir.

$$G(z, X_{maske}, X_{arkaplan}, E(X_{saç})) = X' \qquad \textbf{6.1}$$

Geliştirilen yöntemin stil yeniden yapılandırma görevindeki bazı sonuçları Şekil 6.3'te gösterilmektedir.

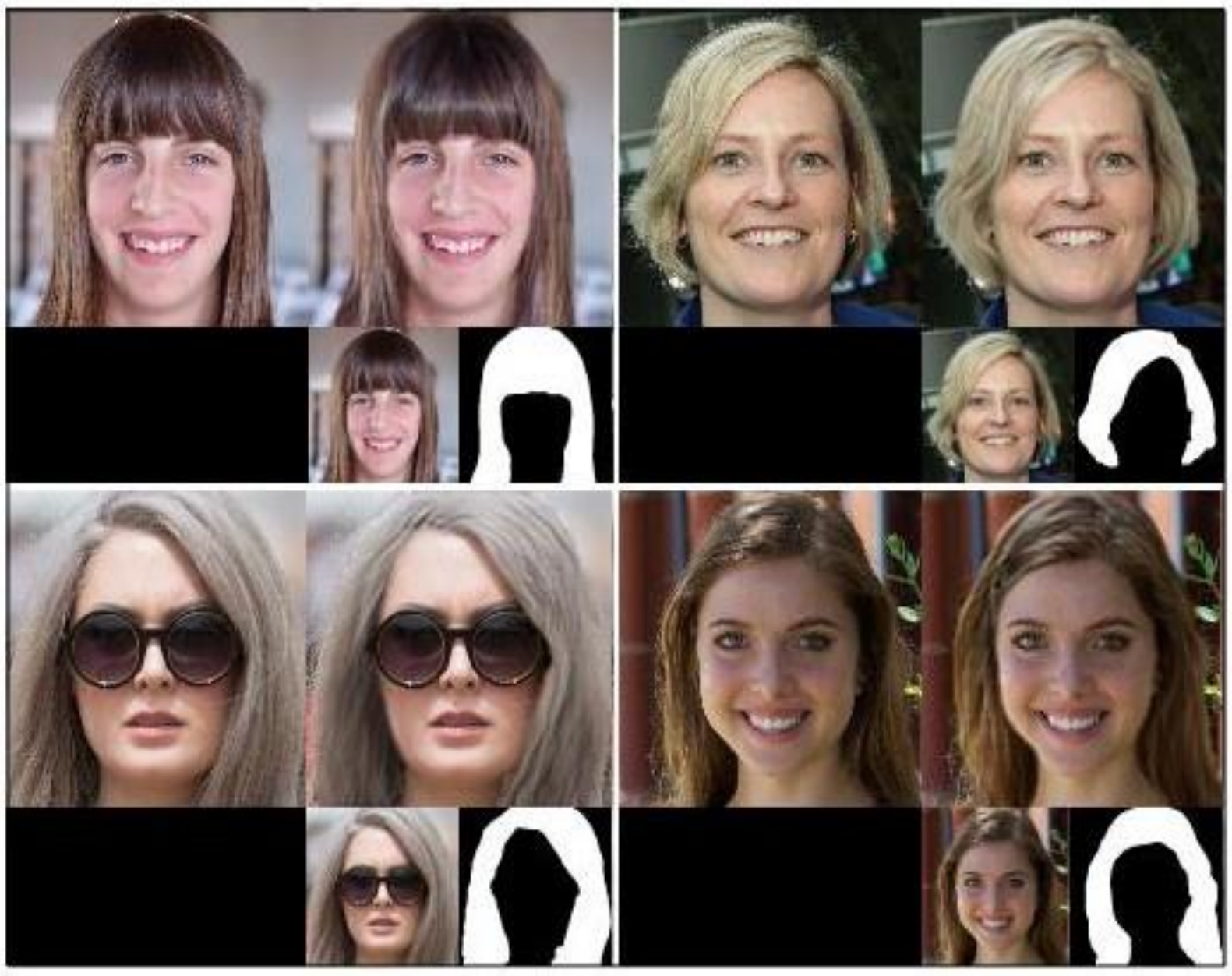

Şekil 6.3. Önerilen yöntemin saç stili yeniden yapılandırma görevi için elde ettiği bazı sonuçlar. Orjinal görüntü sol, sentezlenen görüntü sağ tarafta yer almaktadır.

### 6.3.2 Saç stil transferi

Saç stil transferi görevinde, seçilen bir girdi görseli için farklı bir referans görseli seçilir. Girdi ve referans görselden elde edilen ilgili materyaller yönteme girdi olarak verilir ve sonuç olarak orijinal görüntünün, referans saç stili ile üretilmesi amaçlanır. Saç stili transfer işlemi denklem 6.2 ile formülize edilebilir.

$$G(z, X_{maske}, X_{arkaplan}, E(R_{saç})) = X' \qquad \textbf{6.2}$$

Önerilen yöntem ile elde edilen saç stili transferi sonuçları Şekil 6.4'te gösterilmektedir.

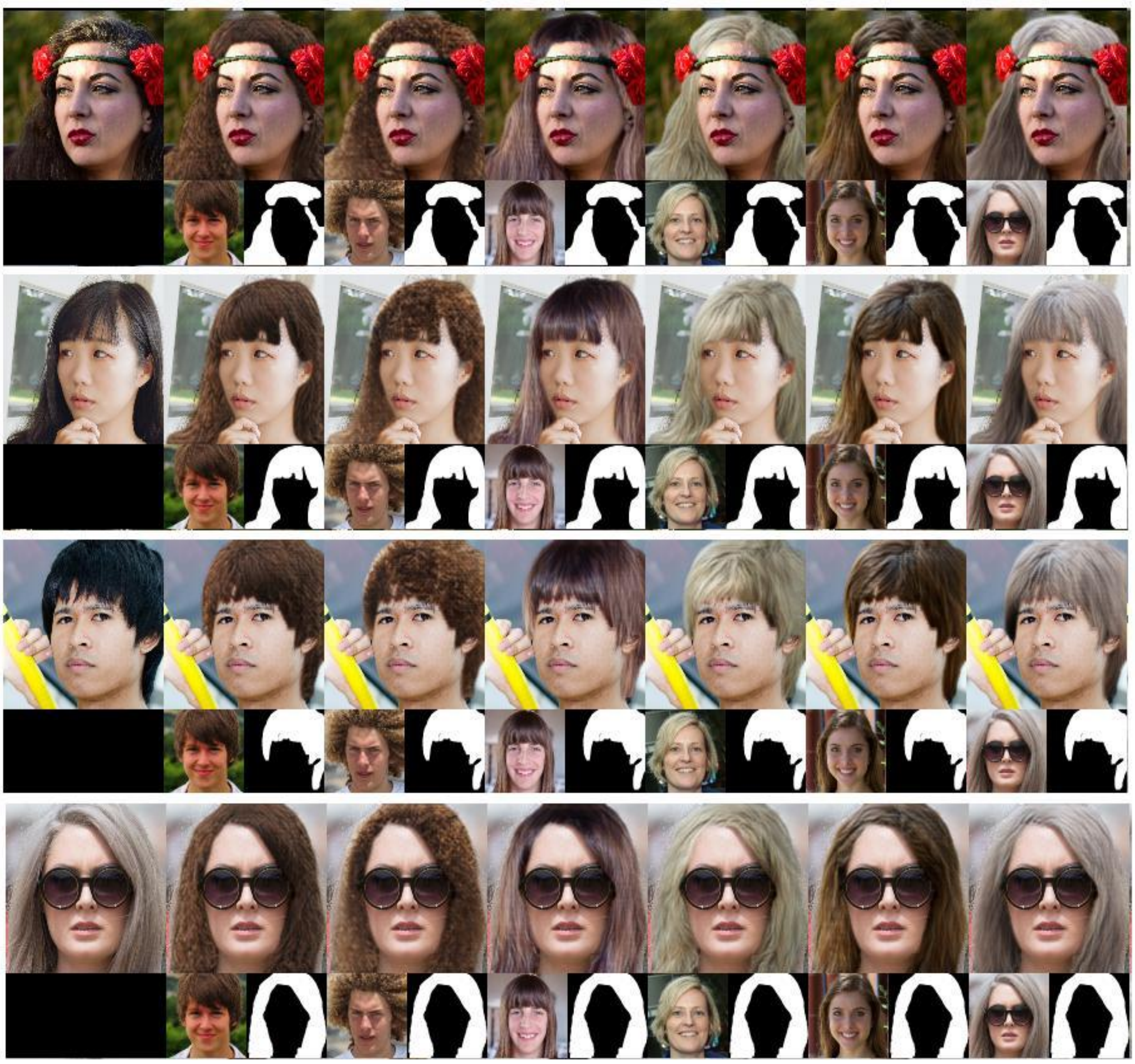

Şekil 6.4. Önerilen yöntemin saç stili transferi görevi için elde ettiği bazı sonuçlar. Orjinal görüntü 1. sütun, sentezlenen görüntüler diğer sütunlarda yer lmaktadır. Referans saç stili görseli ve hedef saç maskesi çıktı görselinin alt kısımda yeralmaktadır.

### 6.3.3 Saç şekli düzenleme

Saç şekli düzenleme görevi, görseldeki saç bölgesinin düzenlenerek farklı şekillere sahip saç üretimini içermektedir. Saçın stilini değiştirmenin yanı sıra, saç şeklinde yapılabilen değişimler daha ileri bir saç sentezme seçeneği sunmaktadır. Bu görev kapsamında, orijinal görsel, referans görsel ve bu görsellere karşılık gelen saç maskeleri elde edildikten sonra orijinal görselin saç maskesi üzerinde bazı düzenlemeler yapılır. Bu düzenlemeler sonucunda elde edilen yeni saç maskesindeki saç bölgelerinde görsele uygun bir şekilde saç üretilmesi hedeflenir. Saç şekli düzenleme görevi denklem 6.3 ile ifade edilebilir.

$$G(z, X_{düzenlenmiş\ maske}, X_{arkaplan}, E(R_{saç})) = X' \qquad \textbf{6.3}$$

Önerilen yöntemin saç şekli düzenleme görevindeki sonuçları Şekil 6.5'te gösterilmektedir.

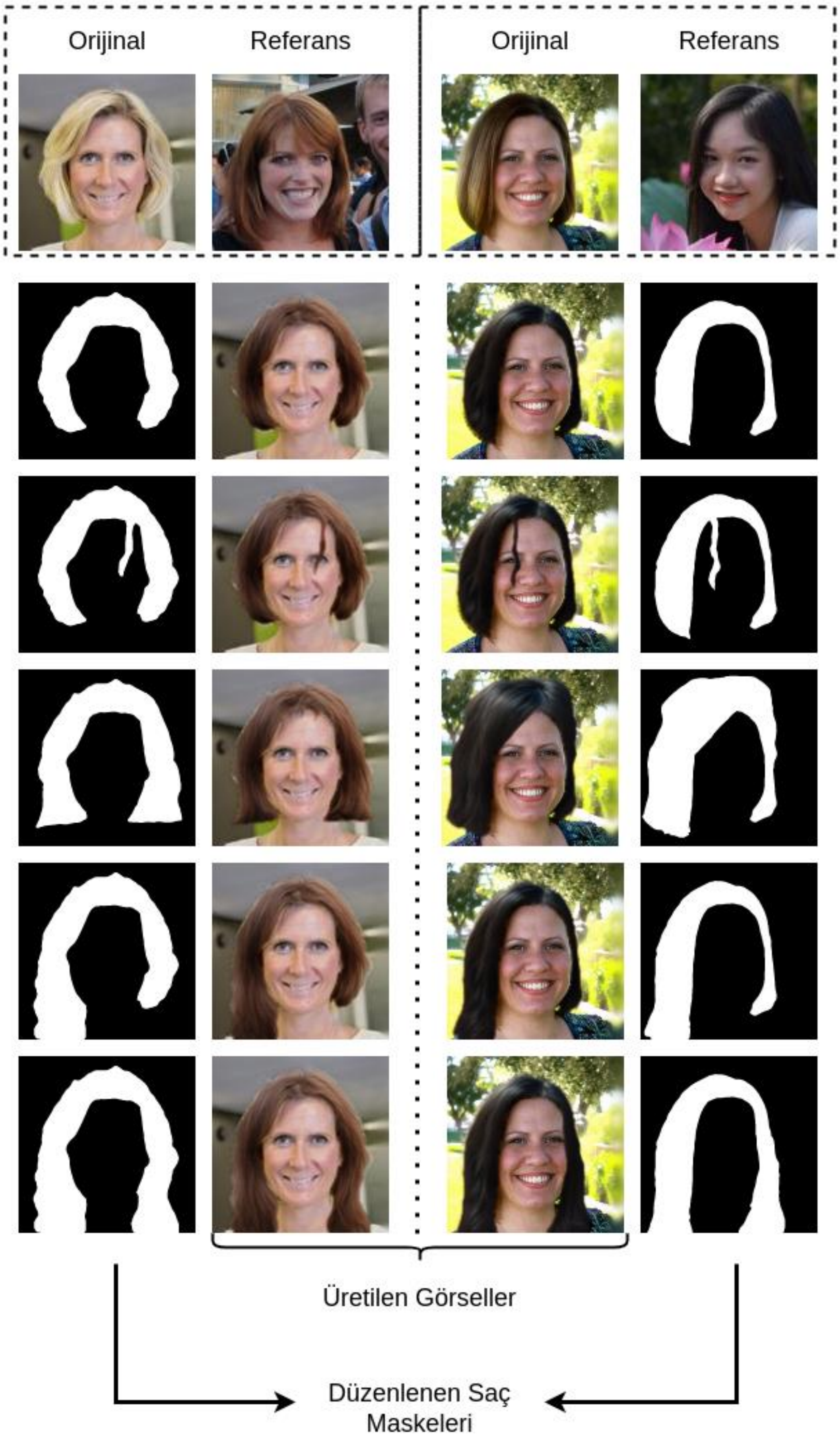

Şekil 6.5 Önerilen yöntemin şaç şekli düzenleme görevinde elde ettiği bazı sonuçlar.

Şekil 6.5'te görüldüğü üzere önerilen yöntem, stil transferi görevi ile birlikte şekilde düzenleme görevinde de başarılı sonuçlar elde etmiştir.

### 6.3.4 Önerilen yöntemin literatür ile karşılaştırılması

Bu bölümde önerilen yöntem, literatürdeki en başarılı sonuçlar seviyesinde sonuçlar elde edebilen MichiGAN yöntemi ile hem üretilen görsellerin kalitesi hem de çalışma zamanı verimliliği açısından değerlendirilmiştir. Önerilen yöntemin ve MichiGAN yönteminin FFHQ veri kümesinden rastgele seçilmiş görseller üzerinde saç yeniden yapılandırma ve saç stil transferi görevleri gerçekleştirilmiş buradan elde edilen çıktılar üzerinden görsel kalitelerinin nicel ve nitel değerlendirilmeleri yapılmıştır.

Saç stil yeniden yapılandırılması görevinde ilgili girdilerden orjinal görselin kendisi sentezlenmesi amaçlanır. Şekil 6.6 incelendiğinde önerilen yöntemin stil yeniden yapılandırma görevide nitel açıdan başarılı çıktılar elde ettiği görülmektedir.

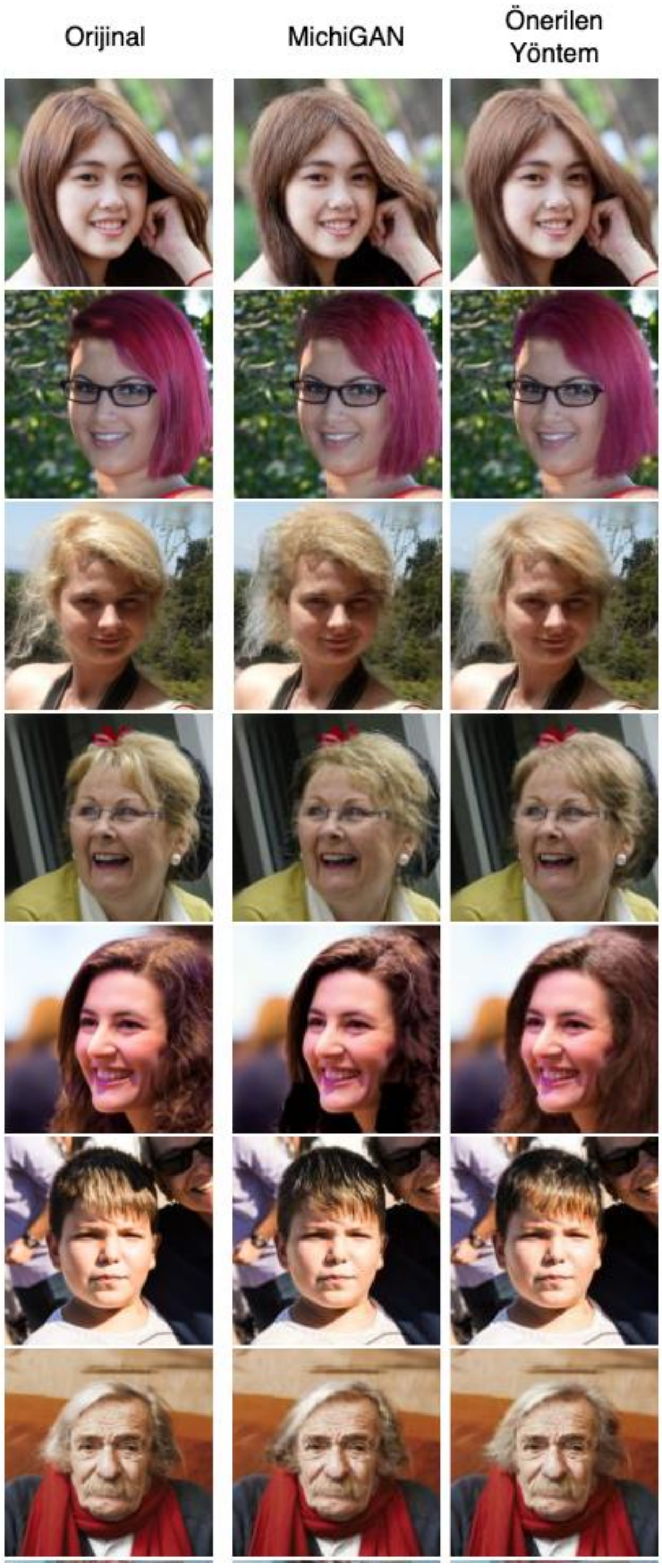

Şekil 6.6. Önerilen yöntemin saç stil yeniden yapılandırması görevi için elde ettiği bazı sonuçlar. Orjinal görüntü ilk sütun, MichiGAN ikinci sütun ve önerilen yöntem üçüncü sütunda gösterilmektedir.

Şekil 6.6 incelendiğinde saç dışındaki bölgenin tamamen korunduğu ve sadece ilgili saç bölgesinin sentezlendiği görülmektedir. Elde edilen sentezlenmiş görüntülerin orjinal görüntüler ile arasındaki farkın az olması, saç stili yeniden yapılandırma görevi için iki yöntemin de yüksek başarılara sahip olduğunu göstermektedir.

Saç stil transferi görevinde, girdi resmine ek olarak farklı bir referans resim seçilir. Seçilen referans resimdeki saç'ın stil özellikleri girdi resmine aktarılması hedeflenir. Bu görevin sonucunda saç dışında bir bölgenin değişmemesi ve saç bölgesi ile arka plan görseli arasındaki geçişin gerçekçi yapılması beklenir. Şekil 6.7'da, önerilen yöntemin literatürdeki en başarılı sonuçlar seviyesinde çıktılar elde edebilen MichiGAN ile karşılaştırması gösterilmektedir.

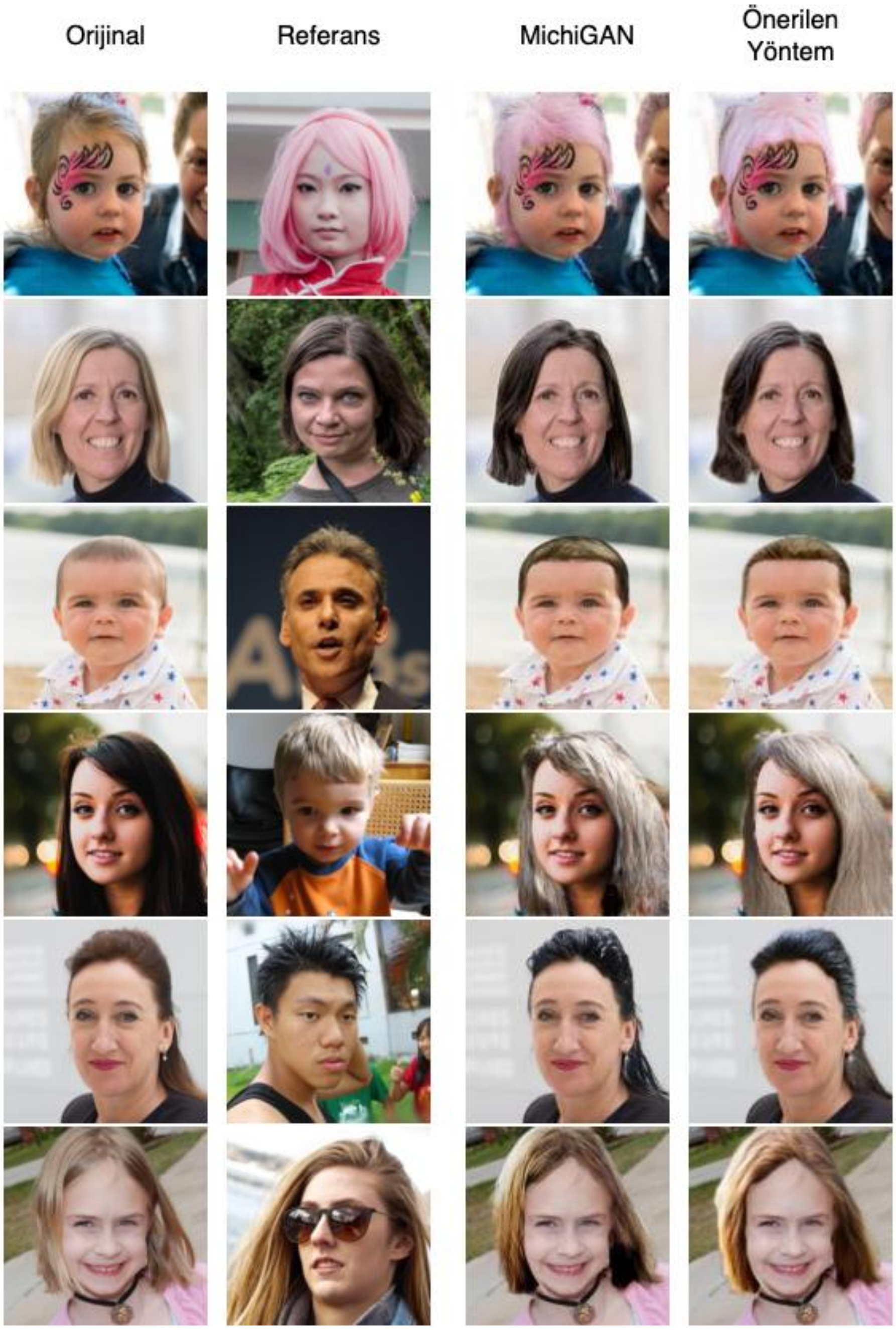

Şekil 6.7. Önerilen yöntemin saç stili transferi görevi için elde ettiği bazı sonuçlar. Orjinal örüntü ilk sütun, referans görüntü ikinci sütun, MichiGAN sonuçları üçüncü sütun ve önerilen yöntemin sonuçları dördüncü sütunda gösterilmektedir.

Yöntemlerin çıktılarının başarısı görsel olarak araştırmacılar tarafından değerlendirilmekle birlikte görsel veri üretiminde bu değerlendirme için sıkça kullanılan metrikler bulunmaktadır. Bu metriklerden en yaygın kullanılanı Frechet Inception Distance (FID)'dir. Önerilen yöntemin MichiGAN ile FID, PSNR ve SSIM metrikleri kullanılarak karşılaştırılması Tablo 6.1'de gösterilmektedir.

Tablo 6.1. Önerilen yöntemin MichiGAN ile FID metriği kullanılarak karşılaştırılması. FID değeri küçüldükçe daha iyi, PSNR ve SSIM metriklerinde ise değer yükseldikçe daha iyi anlamına gelmektedir.

| | **Saç Stil Yeniden Yapılandırma** | | | **Saç Stil Transferi** | | |
|---|---|---|---|---|---|---|
| **Yöntemler** | **FID** | **PSNR** | **SSIM** | **FID** | **PSNR** | **SSIM** |
| MichiGAN | 26.607 | 32.993 | 0.872 | **31.160** | 32.361 | 0.791 |
| Önerilen Yöntem | **26.171** | **35.881** | **0.907** | 33.111 | **34.283** | **0.819** |

Her iki yöntemin sonuçları, FFHQ test veri kümesinden rastgele seçilen 5000 adet orijinal ve referans görsel çiftinden oluşturulan veri kümesi üzerinde değerlendirilmiştir. MichiGAN sonuçları 512x512 çözünürlüğüne sahip olduğundan bilinear interpolasyon yöntemi ile 128x128'e yeniden boyutlandırılmıştır. Sentezlenecek veya stil vektörü ile temsil edilecek bölge saç bölgesi olduğundan, görseller en az %3 saç bölgesi içeren görseller arasından seçilmiştir. Tablo 6.1'de görüldüğü üzere, önerilen yöntem saç yeniden yapılandırma görevinde MichiGAN'dan daha yüksek performans sergilemekle birlikte saç stil transferi görevinde gerisinde kalmıştır. Bu değerlendirmeden de görüldüğü üzere, önerilen yöntem literatürdeki en iyi yöntemler ile yarışır seviyede başarılı görsel çıktılar üretebilmektedir.

Önerilen yöntem görsel başarı olarak literatürdeki en iyi yöntemler başarısında çıktılar üretmeyi hedeflemekle birlikte en önemli hedeflerinden birisi de gerçek zamana yakın hızlarda bu saç sentezi yapabilmektir. Bu bağlamda önerilen yöntemin çalışma hızı ve bellek verimliliği önem kazanmaktadır. Önerilen yöntemin verimlilik açısından başarılı olabilmek için 128x128 çözünürlüğünde görseller üzerinde eğitimi yapılmıştır. Bu durum yüksek çözünürlüklü görseller üretme açısından rakiplerine karşı doğal bir dezavantaja sahip olması anlamına gelirken bellek ve çalışma hızı açısından kendisine avantaj sağlamaktadır. Tablo 6.2'de, önerilen yöntem ile MichiGAN yöntemi çalışma zamanı hızı, parametre sayısı ve görsel çözünürlüğü açısından karşılaştırılmaktadır.

Tablo 6.2. Önerilen yöntemin MichiGAN ile çalışma zamanı hızı, bellek ve görsel çözünürlüğü açısından karşılaştırılması. FPS değeri, 500 adet görsel üzerinden hesaplanmıştır.

| **Yöntemler** | **Görsel / Saniye (FPS)** | **Parametre Sayısı (M)** | **Çözünürlük (Piksel)** |
|---|---|---|---|
| MichiGAN | 2.46 | 109.5 | **512x512** |
| Önerilen Yöntem | **31.25** | **33.12** | 128x128 |

Tablo 6.2'deki sonuçlar bir görsel üretimi için ortalama süre hesaplanılarak elde edilmiştir. Üretim sürecine model yükleme hariç diğer kaynakların yüklenmesi ve kaydedilmesi dahil edilmiştir. Geliştirilen yöntem çıktı çözünürlüğü açısından MichiGAN'dan 4 kat daha düşük bir çözünürlüğe sahip olmasının hız avantajına sahip olduğu da göz önünde bulundurarak MichiGAN'dan yaklaşık 12 kat (31.25/2.46 ~ 12) daha yüksek çalışma hızına ve yaklaşık 3 kat (109.5/33.12 ~ 3) daha düşük bellek alanına sahiptir. Çalışma zamanı hızını ölçebilmek için saniye başına işlenen görüntü sayısı dikkate alınmıştır. Parametre sayıları iki yöntemin de üretici ağları üzerinden hesaplanmıştır.

# 7 SONUÇ VE GELECEK ÇALIŞMALAR

Bu tez çalışmasında saç sentezi probleminin çözümü için bir Çekişmeli üretici ağ mimarisi önerilmiştir. Önerilen yöntemin başarısı saç stil transferi, saç yeniden yapılandırma ve saç şekli düzenleme görevleri için test edilmiştir. Saç stil transferi ve saç yeniden yapılandırma görevlerinde literatürdeki en iyi yöntemlerden birisi olan MichiGAN ile hem nitel hemde nicel karşılaştırmalar yapılmıştır.

Önerilen yöntem bir üretici, bir kodlayıcı ve bir üretici ağ'dan oluşmaktadır. Üretilen saç bölgesinin orjinal görüntü ile başarılı bir şekilde karıştırılması için Saç Karıştırma Katmanı tasarlanmıştır. Bir kodlayıcı ağ vasıtasıyla referans saç'a ait stil vektörü elde edilmiş ve bu stil vektörü AdaIN vasıtasıyla üretici ağ'a enjekte edilmiştir. Bu işlemler sonucunda uçtan uca saç sentezi yapabilen bir yöntem önerilmiştir. Önerilen yöntem, FFHQ veri kümesi ile sözde-denetimli bir şekilde eğitilmiştir.

Eğitim sonucunda elde edilen modelin gerçekçi saç üretimi başarısı FID, PSNR ve SSIM metriği kullanılarak MichiGAN yöntemi ile karşılaştırılmış ve iki yöntem benzer seviyelerde başarılar elde etmişlerdir. Önerilen yöntem, gerçekçi sentezleme yeteneği ile birlikte çalışma zamanı bellek açısından da verimli olmayı da amaçlamaktadır. MichiGAN ile yapılan çalışma zamanı hızı karşılaştırmasında yaklaşık 12 kat daha hızlı ve yaklaşık 3 kat daha az bellek alanına sahip olduğu gösterilmiştir. Bu hız değerlendirilirken MichiGAN'ın 512x512 çözünürlükte görseller ürettiği ancak önerilen yöntemin 128x128 çözünürlüğe sahip görseller ürettiği göz önünde bulundurulmalıdır.

Sonuç olarak önerilen yöntem literatürdeki en iyi yöntemlerle yarışır kalitede saç sentezi yapabilirken, gerçek zamanlı bir işletim hızına da sahiptir. Yapılan çalışmanın yakın gelecekte, araştırmacılar için gerçek zamanlı saç sentezi çalışmalarına katkı sağlayacağını ön görülmektedir.

# KAYNAKLAR DİZİNİ

## KAYNAKLAR DİZİNİ (devam)

## KAYNAKLAR DİZİNİ (devam)

## KAYNAKLAR DİZİNİ (devam)

## KAYNAKLAR DİZİNİ (devam)

## KAYNAKLAR DİZİNİ (devam)

## KAYNAKLAR DİZİNİ (devam)

## KAYNAKLAR DİZİNİ (devam)

## KAYNAKLAR DİZİNİ (devam)

# TEŞEKKÜR

Tez çalışmam boyunca desteğini eksik etmeyen nişanlım Cevher SÖYLEMEZ'e, çalışma boyunca bana özgür bir çalışma ortamı sağlayan ve destekleyen danışmanım Prof. Dr. Aybars UĞUR'a ve hem donanım ihtiyaçlarım hemde bünyesinde çalışırken yüksek lisans eğitimime devam edebilmem konusunda gerekli desteği sağlayan Syntonym A.Ş'ye çok teşekkür ederim.

Tabiki eğitimimin en başından beri beni destekleyen başta babam Cemil PEKTAŞ ve annem Muteber PEKTAŞ olmak üzere tüm aileme minnet ve teşekkürlerimi sunarım.

28 / 01 / 2022

Muhammed PEKTAŞ

# ÖZGEÇMİŞ

**Kişisel Bilgiler**

Soyadı, adı : PEKTAŞ, Muhammed

Uyruğu : Türkiye

**Eğitim**

| Derece | Eğitim Birimi | Mezuniyet tarihi |
|---|---|---|
| Yüksek Lisans | Ege Üni./ Bilgisayar Mühendisliği | 2022 |
| Lisans | Konya Teknik Üni./Bilgisayar Mühendisliği | 2019 |

**İş Deneyimi**

| Yıl | Yer | Görev |
|---|---|---|
| 2019 | Syntonym A.Ş | Bilgisayarlı Görü Mühendisi |
| 2018 | IBTECH | Stajyer |
| 2017 | Microsoft Turkiye | Stajyer |

# EKLER

Ek 1. İngilizce-Türkçe Sözlük

## Ek 1. İngilizce-Türkçe Sözlük

| **İngilizce** | **Türkçe** |
|---|---|
| Machine Learning | Makine Öğrenmesi |
| Deep Learning | Derin Öğrenme |
| Virtual Reality | Sanal Gerçeklik |
| Augmented Reality | Artırılmış Gerçeklik |
| Bias | Yanlılık |
| Classification | Sınıflandırma |
| Dataset | Veri seti |
| Noise | Gürültü |
| Generative Adversarial Networks | Çekişmeli Üretici Ağlar |
| Loss Function | Kayıp Fonksiyonu |
| Gradient Descent | Gradyan İniş |
| Ground Truth | Gerçek Değer |
| Adaptive Instance Normalization | Adaptif Örnek Normalleştirmesi |
| Psuedo Supervised Training | Sözde Denetimli Öğrenme |